\documentclass[a4paper,11pt]{article}
\usepackage{jheppub}
\usepackage[T1]{fontenc}
\usepackage{graphicx}
\usepackage{amsmath}
\usepackage{amsmath}
\usepackage{xspace}
\usepackage{subfig}
\usepackage{color}
\usepackage[utf8]{inputenc}
\usepackage{commath}
\usepackage{slashed}
\usepackage{fancyvrb}

\title{\boldmath Displaced fat-jets and tracks to probe boosted right-handed neutrinos in the $U(1)_{B-L}$ model}
\newcommand{\AddrHBNI}{
	Homi Bhabha National Institute, BARC Training School Complex, Anushakti Nagar, Mumbai 400094, India }

\author[a,b]{Rojalin Padhan,}
\author[a,b]{Manimala Mitra,}
\author[c]{Suchita Kulkarni,}
\author[d]{Frank F. Deppisch}
\affiliation[a]{Institute of Physics, Sachivalaya Marg, Bhubaneswar, Odisha 751005, India}
\affiliation[b]{\AddrHBNI}
\affiliation[c]{Institute of Physics, NAWI Graz, University of Graz,Universit\"atsplatz 5, A-8010 Graz, Austria}
\affiliation[d]{University College London, Gower Street, London WC1E 6BT, UK}

\emailAdd{rojalin.p@iopb.res.in}
\emailAdd{manimala@iopb.res.in}
\emailAdd{suchita.kulkarni@uni-graz.at}
\emailAdd{f.deppisch@ucl.ac.uk}
\preprint{IP/BBSR/2022-02}

\abstract{We investigate the pair-production of Right-Handed Neutrinos (RHNs) via a $B-L$ $Z'$ boson and the detection prospects at the High-Luminosity run of the LHC (HL-LHC) and a future $pp$ collider (FCC-hh). We focus on RHN states with a mass of $10-70$~GeV which naturally results in displaced vertices for small active-sterile mixing strengths. Being produced through a mass resonance with $m_{Z'} \ge 1$~TeV, the RHNs are heavily boosted, leading to collimated decay products that give rise to fat-jets. We investigate the detection prospect of dedicated signatures in the inner detector and the muon spectrometer, namely a pair of displaced fat-jets and the associated tracks, respectively. We find that both the HL-LHC and FCC-hh can be sensitive to small active-sterile mixing $V_{\mu N} > 10^{-6}$ and $V_{\mu N} > 10^{-7}$ with the number of events reaching $\mathcal{O}(10)$ and $\mathcal{O}(10^3)$, respectively. This allows probing the generation of light neutrino masses through the Seesaw mechanism in this scenario.}

\begin{document} 
\maketitle
\flushbottom

%%%%%%%%%%%%%%%%%%%%%%%%%%%%%%%%%%%%%%%%%%%%%%%%%%%%%%%%%%%%%%%%%%%%%%%%%%%%%%%%%
\section{Introduction}
%%%%%%%%%%%%%%%%%%%%%%%%%%%%%%%%%%%%%%%%%%%%%%%%%%%%%%%%%%%%%%%%%%%%%%%%%%%%%%%%%
The discovery of the Standard Model (SM) like  Higgs boson at the Large Hadron Collider (LHC) experiment has  proven that in the SM, masses of the fermions and gauge bosons  are generated via the Brout-Englert-Higgs (BEH) mechanism. However, the origin of light neutrino masses and mixing still remains a key question, which can not be  explained by the SM.  A number of neutrino oscillation experiments have  observed that, the solar and atmospheric neutrino mass splittings are $\Delta m^2_{12} \sim 10^{-5}$ $ \rm{eV}^2$ and $\Delta m^2_{13} \sim 10^{-3}$ $ \rm{eV}^2$, and the mixing angles are $\theta_{12} \sim 32^\circ$, $\theta_{23} \sim 45^\circ$, and 
$\theta_{13} \sim 9^\circ$  \cite{Esteban:2020cvm}. There have been a number of  proposed beyond the Standard Model (BSM) extensions, which can explain the observed light neutrino masses and mixings and  contain SM gauge singlet right-handed neutrinos (RHNs). The $U(1)_{B-L}$ model is such an extension, where the gauge sector is also extended by an additional $U(1)_{B-L}$ gauge group \cite{Mohapatra:1980qe, Wetterich:1981bx, Georgi:1981pg,Davidson:1978pm}. The model contains three right handed neutrinos $N_i$, which are SM gauge singlet, however carry non-trivial  charges under the $U(1)_{B-L}$ gauge symmetry. Other than the three RHNs, the model also contains an extended gauge sector, and a complex scalar $S$. The SM neutrino mass in this model is generated from the  lepton number violating (LNV) $d=5$ seesaw operator $LLHH/\Lambda$~\cite{Weinberg:1979sa,Wilczek:1979hc} via Type-I seesaw mechanism 
\cite{Minkowski:1977sc,Mohapatra:1979ia,Yanagida:1979as,GellMann:1980vs,Schechter:1980gr}.  

{Being SM gauge singlet RHNs interact with the SM particles only via their mixing with the active neutrinos, referred as active-sterile mixing, which  is proportional to $\sqrt{{m_\nu}/{M_N}}$, where $m_{\nu} $ and $M_N$ are the light neutrino and the RHN mass scales, respectively.  Since  $m_\nu <$ eV, this mixing is small for RHN with mass few tens to hundreads of GeV, leading to an in-general  suppressed  production of a single RHN at a $pp$ collider.  This limitation can however be evaded in the gauged $B-L$ model, since RHNs can be produced via  additional  production mechanism that involves unsuppressed interactions of RHN with  BSM/SM particles~\cite{PhysRevLett.44.1316,WETTERICH1981343}.} 

{Motivated by this, we study pair-production of RHNs in gauged $B-L$ model, where the production primarily occurs via mediation of BSM gauge boson $Z^{\prime}$. Specifically, we focus on the low mass region of $N$ satisfying $M_N< M_W$, for which $N$ gives rise to distinctive model signatures. The produced $N$ undergoes three body decays  to few different final states, among which we consider $N \to \mu q q^{\prime}$ decay mode. For the considered mass range, two body decay of $N$ is kinematically forbidden. Since the decay of $N$ strongly  depends on the active-sterile mixing, hence for mixing in agreement with eV light neutrino mass, the RHN undergoes   displaced decays where its decay vertex is considerably displaced from the production vertex. Additionally, due to a large hierarchy between the masses of the $Z^{\prime}$ and $N$, the decay product of $N$ is collimated, resulting in a displaced  {\it fat-jet} signature.  We hence study distinctive long-lived signatures of $N$, that can be detected at the  High-Luminosity upgrade  of the LHC,  and the future FCC-hh machine. For previous studies on displaced RHN decay, see \cite{Deppisch:2018eth,Antusch:2017hhu,Cottin:2018nms,Drewes:2019fou,Abada:2018sfh,Liu:2019ayx,Liu:2022kid}.}

{In particular, we consider  displaced decays of $N$ in two different region of the detectors, a)  in the inner detector (ID) and b) the farthest end of the detector, which is muon spectrometer~(MS). To realise the former,  $N$ must  have proper decay length ranging between  $\ell \sim$ mm to hundreds of mm, while for later, the decay length should be $\ell \sim $ m.  We note   that the  signal description between a) and b) differ widely.
	For the decay of $N$ in ID,  the signature contains two displaced {\it fat-jets}. For $N$ decaying in the MS, the  track properties  in the ID, energy deposits in the calorimeter can not be used, which are used for the jet formation. Therefore, a  jet description of the final state particles is inadequate in this case, and we instead perform  a track-based analysis.  We further extend the analysis  by demanding displaced decay of at-least one RHN in the ID/MS.} {Assuming a background free environment we find  that few tens of  displaced {\it fat-jet} events can be observed at the  HL-LHC for c.m.energy $\sqrt{s}=14$ TeV and luminosity $\mathcal{L}=3\, \rm{ab}^{-1}$. For FCC-hh that can operate with a much higher c.m.energy $\sqrt{s}=100$ TeV and luminosity $\mathcal{L}=30\, \rm{ab}^{-1}$ , the maximum achievable events  increases by order of magnitude. For the decay of $N$ in the MS, the discovery prospect at the HL-LHC is rather low. This however improves significantly for FCC-hh.}

The discussion proceeds as follows: In Section.~\ref{blreview}, we present a brief review of the model, following which we present a discussion on existing constraints in Section.~\ref{expcons}.   In Section.~\ref{sec:HPD}, and Section.~\ref{indmuspec},  we discuss the displaced decays and pair-production of two $N$s. We present an extensive analysis for HL-LHC and FCC-hh in Section.~\ref{hllhcproj} and Section.~\ref{fcchhproj}, respectively. Finally, we present a summary in Section.~\ref{conclu}.
%%%%%%%%%%%%%%%%%%%%%%%%%%%%%%%%%%%%%%%%%%%%%%%%%%%%%%%%%%%%%%%%%%%%%%%%%%%%%%%%%
\section{The $B-L$ gauge model}
\label{blreview}
%%%%%%%%%%%%%%%%%%%%%%%%%%%%%%%%%%%%%%%%%%%%%%%%%%%%%%%%%%%%%%%%%%%%%%%%%%%%%%%%%
%\subsection{Model setup and particle spectrum}
%%%%%%%%%%%%%%%%%%%%%%%%%%%%%%%%%%%%%%%%%%%%%%%%%%%%%%%%%%%%%%%%%%%%%%%%%%%%%%%%%
We consider the gauge $B-L$ model,  which  in addition to the particles of the SM sector also contains three right-handed neutrinos $\nu_R$, a BSM Higgs field $S$ and a BSM gauge boson $B^{\prime}_{\mu}$. The gauge group is $SU(3)_c\times SU(2)_L \times U(1)_Y \times U(1)_{B-L}$, where the scalar and fermionic fields $S$ and ${\nu_R}_i$ have $B-L$ charges $B-L = +2$ and $-1$, respectively. 
%The charge assignments of different particles are shown in Table~\ref{tabbml}.
The SM states have the conventional $B-L$ charges.
 Here, the field $S$ represents a complex scalar field, which acquires vacuum expectation value~(vev) $v_{BL}\neq 0$ and breaks the $B-L$ gauge symmetry. The states ${\nu_R}_i$ contribute to the light neutrino mass generation via the seesaw mechanism. The complete Lagrangian of the model has the form  
%\begin{table}[h]
%	\begin{tabular}{ |c| c |c |c| c|c |c|c |c|}
%		\hline 
%		& $\phi$ & $\nu_R$ &  $L$ & $Q$ & $u_R$ & $d_R$ & $e_R$ & $S$\\ \hline
%		$Y_{B-L}$ & $0$ & $-1$ & $-1$ & $1/3$ &$1/3$ & $1/3$ &$1$& $2$ \\
%		\hline
%	\end{tabular}
%	\caption{Charges of all the particles under $B-L$  symmetry.}  \label{tabbml}
%\end{table}
       
	\begin{equation}
	\mathcal{L}= \mathcal{L}_{SM}+\mathcal{L}_{B-L},
	\end{equation}
	where  $\mathcal{L}_{B-L}$ is the $B-L$ Lagrangian, and $\mathcal{L}_{SM}$ is the  Lagrangian for the SM sector.  The $B-L$ Lagrangian  has the form
	{\small\begin{equation}
	\begin{aligned}
	\mathcal{L}_{B-L}=&\left(D_{\mu} {S}\right)^{\dagger}\left(D^{\mu} {S}\right)-\frac{1}{4} F_{B L \mu \nu} F_{B L}^{\mu \nu}+{i} {{\bar{\nu}}_{Ri}} \gamma^{\mu} D_{\mu} {\nu_R}_{i}-V_{B-L} \left({\phi}, {S}\right)\\
	&-\sum_{i=1}^{3} y^{M}  {S} \bar{\nu_R}_{i}^{c} {\nu_R}_{i}-\sum_{i, j=1}^{3} y^{\nu}_{ij} \bar{L}_{i} \tilde{\phi} {\nu_R}_{j}+h . c . \  ,
	\end{aligned}
	\label{eq:LBL}
	\end{equation}}
with {\small\begin{equation}
	V_{B-L}\left(\phi, S\right)=\mu_{S}^{2} S^{\dagger} {S}+\mu_{\phi}^{2} {\phi}^{\dagger} {\phi}+\lambda_{1}\left({\phi}^{\dagger} {\phi}\right)^{2}+\lambda_{2}\left({S}^{\dagger} {S}\right)^{2}+\lambda_{3}\left({\phi}^{\dagger} {\phi}\right)\left({S}^{\dagger} {S}\right).
	\end{equation}}\\
	In the above, $D_\mu$ represents the covariant derivative~\cite{Pruna:2011me}, 
%-------------------------------------------------------------------------------------------------------------------------------------------------------------------------------------------------------------------------------
\begin{align}
\label{DM}
	D_\mu = \partial_{\mu} + ig_{s}\mathcal{T}_\alpha G_\mu^\alpha 
	      + igT_a W_\mu^a + ig_1 Y B_\mu + i (\tilde{g}Y + g^{\prime} Y_{B-L}) B^\prime_\mu ,   
\end{align} 
%---------------------------------------------------------------------------------------------------------------------------------------------------------------------------------------------------------------------------------------------------------------------
where $G^\alpha_\mu$, $W^a_\mu$, $B_\mu$ are the  SM gauge fields with associated couplings $g_s$, $g$, $g_1$ and the respective generators are $\mathcal{T}_\alpha$, $T_a$, $Y$, respectively. The field denoted as $B^\prime_\mu$ represents  the gauge field for  $U(1)_{B-L}$ gauge symmetry,  $g^{\prime}$ represents the respective gauge coupling  and the $B-L$ quantum number is denoted by $Y_{B-L}$. In this paper, we neglect the mixing between $U(1)_{B-L}$ and $U(1)_{Y}$ to simplify the model, i.e. we consider the minimal gauged $B-L$ model. This model is therefore valid in the limiting case of small mixing   between $B_\mu$ and $B^\prime_\mu$. The gauge sector of the model also includes the following kinetic term  for the gauge field $B^{\prime}_{\mu}$,
%-------------------
%---------------------------------------------------------------------------------------------------------------------------------------------------------------------------------------------------------------------------------------------------------------------
\begin{align}
\label{LYM}
	{\cal L}^{BL}_{kin} =
	-\frac{1}{4} F^{\prime\mu\nu} F_{\mu\nu}^\prime,
\end{align} 
%---------------------------------------------------------------------------------------------------------------------------------------------------------------------------------------------------------------------------------------------------------------------
where $F^\prime_{\mu\nu} = \partial_\mu B^\prime_\nu - \partial_\nu B^\prime_\mu$ is  the field strength tensor of the $B-L$ gauge group.
The kinetic term for the RHN  and  the SM fermion fields are 
%---------------------------------------------------------------------------------------------------------------------------------------------------------------------------------------------------------------------------------------------------------------------
\begin{align}
\label{Lf}
	{\cal L} =
    i\overline{\nu_{Ri}}\gamma_\mu D^\mu \nu_{Ri}+  i\overline{\psi_{i}}\gamma_\mu D^\mu \psi_{i}
\end{align} 
%---------------------------------------------------------------------------------------------------------------------------------------------------------------------------------------------------------------------------------------------------------------------
In the above, $\psi$ represents the SM fermion fields. These fields receive an additional term in their covariant derivatives, since they are non-trivially charged under the $B-L$ gauge symmetry, with  $Y_{B-L} = 1/3$ and $-1$ for the quark and lepton fields, respectively. In the above a summation over the fermion species and generations is implied.  

\begin{itemize}
	\item {\it Neutrino mass:}  In Eq.~(\ref{eq:LBL}), the Yukawa matrix $y^\nu$  represents the Dirac Yukawa coupling  and $y^M$ is the Yukawa coupling connecting  the RHNs with the complex scalar field $S$. The RHN mass is generated due to breaking of the $B-L$ symmetry, with the mass matrix given by $M_R = y^M \langle S \rangle$. The light neutrinos mix with the RHNs  via the Dirac mass matrix $m_D = y^\nu v/\sqrt{2}$. 
The complete mass matrix in the $(\nu_L, \nu^c_R)$ basis has the form
%%--------------------------------------------------------------------------------
\begin{align}
\label{MD}
	{\cal M} = 
	\begin{pmatrix}
		0   & m_D \\
		m^T_D & M_R
\end{pmatrix},
\end{align} 
%---------------------------------------------------------------------------------------------------------------------------------------------------------------------------------------------------------------------------------------------------------------------
where 
%---------------------------------------------------------------------------------------------------------------------------------------------------------------------------------------------------------------------------------------------------------------------
\begin{align}
\label{MDM}
	m_D = \frac{y^\nu}{\sqrt{2}}v, \quad M_R = \sqrt{2} y^M v_{BL}. 
\end{align} 
Here, $v = \langle \phi^0 \rangle$ and $v_{BL} = \langle S \rangle$ are the vacuum expectation values for electroweak and $B-L$ symmetry breaking, respectively. In the seesaw limit, $M_R \gg m_D$, the light and heavy neutrino masses are 
\begin{align}
\label{seesaw}
	m_\nu \sim - m_D M^{-1}_R m^T_D, \quad M_N \sim M_R.
\end{align}
%---------------------------------------------------------------------------------------------------------------------------------------------------------------------------------------------------------------------------------------------------------------------
The flavour and mass eigenstates of the light and heavy neutrinos are connected as  
%---------------------------------------------------------------------------------------------------------------------------------------------------------------------------------------------------------------------------------------------------------------------
\begin{align}
\label{Neutrino}
	\begin{pmatrix}
		\nu_L \\ \nu^c_R
	\end{pmatrix} = 
	\begin{pmatrix}
		V_{ll} & V_{lN} \\
		V_{Nl} & V_{NN}
	\end{pmatrix}
	\begin{pmatrix}
		\nu^m_L \\ {N^c_R}
	\end{pmatrix},
\end{align} 
where $\nu^m_L$ and $({N^c_R})^T$ represent the left-chiral  mass basis for the light and heavy neutrinos, respectively. In our subsequent discussion, we represent the  physical Majorana fields for the light and heavy neutrinos via $\nu^m=\nu^m_L+(\nu^m_L)^c$ and $N=N_R+(N_R)^c$, respectively. In the above, we schematically  write  the 6-dimensional mixing matrix in terms of 3-dimensional blocks. Assuming the charged lepton mass matrix to be diagonal, the sub-block  $V_{ll} $ can  approximately be considered as the  PMNS mixing matrix $U_\text{PMNS}$. The other sub-block $V_{lN}$ represent the mixing between the light and RHN states and is referred as active-sterile mixing. In general, $V_{lN}$ is an arbitrary $3\times 3$ matrix. However, to pursue a collider study on the proposed signature,  it is sufficient for us to consider one generation of RHNs and only non-zero $V_{\mu N}$, which we follow in the subsequent sections.

		\item {\it $Z^{\prime}$ Gauge boson mass:} Due to the presence of an additional $U(1)_{B-L}$ gauge symmetry, the model contains BSM gauge bosons $B^{\prime}_{\mu}$. We refer to the massive state as $Z^{\prime}$.  Similar to the  RHNs, the additional neutral gauge boson mass $M_{Z^{\prime}}$  is generated via spontaneous  breaking of $B-L$ gauge symmetry. The mass of $Z^{\prime}$ is related to the symmetry breaking scale $v_{BL}$ as  
	\begin{equation}
	M_{Z^{\prime}}=2g^{\prime}v_{BL},
	\end{equation}
	where $g^{\prime}$ is the associated $B-L$ gauge coupling constant.
	
	\item {\it SM Higgs and BSM Higgs:} After spontaneous symmetry breaking (SSB), the SM Higgs doublet $\phi$ and BSM scalar $S$ is given by
	\begin{equation}
	\phi=
	\begin{pmatrix}
	0 \\
	\dfrac{v+h_{1}}{\sqrt{2}}
	\end{pmatrix},
	\ \ \ 
	S=
	\begin{pmatrix}
	\dfrac{v_{BL}+h_{2}}{\sqrt{2}}
	\end{pmatrix},\
	\end{equation}
	with the dynamical states $h_1$ and $h_2$.
	Owing to the non-zero $\lambda_{3}$, $h_1$ and $h_2$ mix with each other which leads to the scalar mass matrix  given by
	\begin{eqnarray}
	\mathcal{M}^2_{\text{scalar}} = \left(\begin{array}{cc}
	2\lambda_1 v^2 ~~&~~ \lambda_{3}\,v_{BL}\,v \\
	~~&~~\\
	\lambda_{3}\,v_{BL}\,v ~~&~~ 2 \lambda_2 v^2_{BL}
	\end{array}\right) \,\,.
	\label{mass-matrix}
	\end{eqnarray}
	The basis states $h_1$ and $h_2$ can be rotated by suitable angle $\alpha$ to the new basis states $H_1$ and $H_2$. The new basis states represents the physical basis states which are given by,
	\begin{eqnarray}
	H_{1}&=&h_{1}\ \cos\alpha - h_{2}\ \sin\alpha,\\
	H_{2}&=&h_{1}\ \sin\alpha + h_{2}\ \cos\alpha,	
	\end{eqnarray}
	where $H_1$ is the SM-like Higgs and $H_2$ is the mostly BSM Higgs. The mixing angle between the two states is
	\begin{equation}
	\tan2\alpha =\frac{v v_{BL}\lambda_{3}}{v^{2}\lambda_{1}-v_{BL}^{2}\lambda_{2}}. 
	\end{equation}
	The mass square eigenvalues of $H_1$ and $H_2$ are given by,
	\begin{equation}
	M_{H_1,H_2}^2=\lambda_1 v^2+\lambda_{2}v_{BL}^2\pm \sqrt{(\lambda_{1} v^2-\lambda_{2}v_{BL}^2)^{2}+(\lambda_{3}vv_{BL})^2}. 
	\label{eq:massesscalar}
	\end{equation}
	\end{itemize}
	In what follows, we consider the scalar mixing to be negligible, for which the SM and BSM  Higgs masses have the form, respectively
	\begin{equation}
	M_{H_1} \sim \sqrt{\lambda_1} v, \, \, \, \,\, 
	M_{H_2} \sim \sqrt{\lambda_2} v_{BL} % \\+\lambda_{2}v_{BL}^2\pm \sqrt{(\lambda_{1} v^2-\lambda_{2}v_{BL}^2)^{2}+(\lambda_{3}vv_{BL})^2}. 
	\label{eq:scalarapprox}
	\end{equation}
	Due to this choice of small mixing, any BSM Higgs production and its decay to two RHNs will be suppressed. 
	
 The RHNs have charged current and neutral current interactions with the SM fields, with the scalars $H_1,H_2$ and the gauge boson $Z^{\prime}$. The respective charged current  interaction Lagrangian  has the form,
	
\begin{align}
-\mathcal{L}_{CC} \  = \ \frac{g}{\sqrt{2}} W^-_{\mu}\bar{\ell} \gamma^{\mu} \nu_\ell + {\rm H.c.} 
\  = \ \frac{g}{\sqrt{2}} W^-_{\mu}\bar{\ell} \gamma^{\mu} \left( V_{ll} P_L {\nu^m_l}+ V_{lN}  P_L {N} \right) + {\rm H.c.},
\label{CC}
\end{align}
where $P_L$ is the left-chirality projection operator $P_L=\frac{1-\gamma^5}{2}$. Similarly, the NC interaction is given by 
\begin{equation}
	\begin{aligned}
-\mathcal{L}_{NC}= &\frac{g}{2 \cos\theta_w}  Z_{\mu}  \bar{\nu_{\ell}} \gamma^\mu  \nu_{\ell} + {g^{\prime}}  Z^{\prime}_{\mu} \bar{\nu}_R \gamma^\mu \nu_R \\
		&\approx   \frac{g}{2 \cos\theta_w}  Z_{\mu} \Big[
		{\bar{\nu}^m}_i \gamma^{\mu} P_L {\nu^m_i}  
		+  \{(V_{ll}^{\dagger} V_{lN})_{ij} {\bar{\nu}}_i \gamma^{\mu} P_L  {N}_j 
		+ {\rm H.C.}\}\Big]	+ g^{\prime}  Z^{\prime}_{\mu}  {\bar{N}}_i \gamma^{\mu} P_R {N}_i,
		\label{NC}
	\end{aligned}
\end{equation}
In the above, $i,j$ represent generation indices.
% and the analytic form of  $\kappa$  is given in the appendix \ref{ }. 
The interaction with the SM and BSM Higgs have the form
   \begin{equation}
	\mathcal{L}^N_{int}=y^M \cos\alpha \ \bar{N} H_2 N+ y^M \sin\alpha \ \bar{N} H_1 N+\bigg[V_{ll}^{\dagger} V_{lN}\cos\alpha \ \bar{\nu^m} H_1 N+H.C. \bigg].
	   \label{eq:scal}
  \end{equation}
%------------------------------------------------------------------------------------------%---------------------------------------------------------------------------------
 \section{Experimental constraints \label{expcons}}
 \begin{figure}
 	\centering
 	\includegraphics[width=0.5\textwidth,height=0.32\textheight]{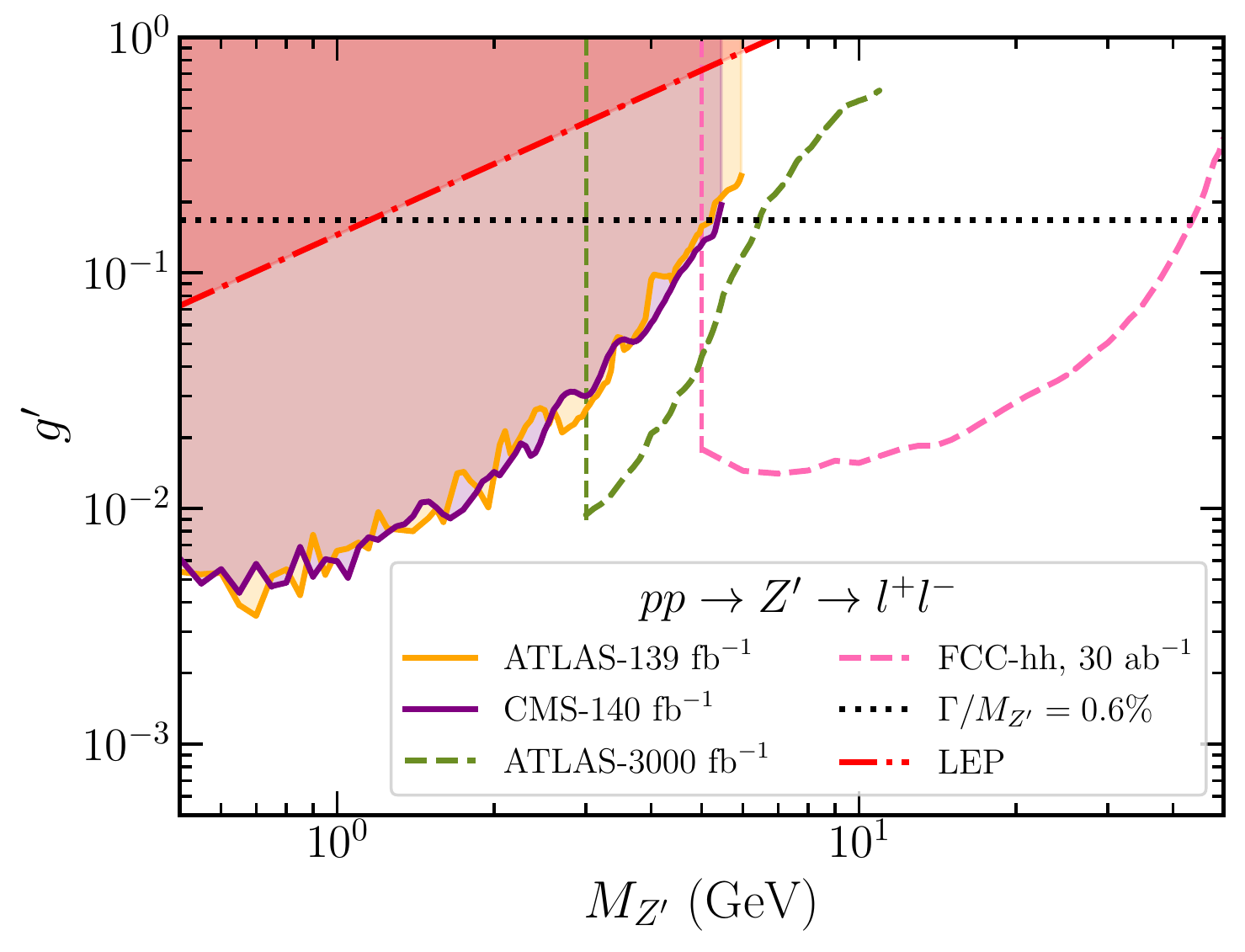}
 	\caption{Limits in the $M_{Z^{\prime}}$ and $g^{\prime}$ plane derived from ATLAS~\cite{ATLAS:2019erb}, and CMS~\cite{CMS:2021ctt} searches. The limit from LEP-II \cite{SLACE158:2003onx} has been shown by the dot-dashed red line. The green dashed and orange dashed  lines represent the projection  for HL-LHC \cite{Ruhr:2016xsg} and FCC-hh~\cite{Helsens:2019bfw}, respectively. {The  black horizontal dotted line represents $\Gamma/M_{Z^{\prime}}=0.6\%$.  }}
 	\label{fig:lhclimit}
 \end{figure}
 The $Z^{\prime}$ and $N$  of the gauged $B-L$ model are significantly  constrained from the non-observation of any direct signature at LEP, and  LHC. In addition, eV light neutrino masses impose additional constraints on the model parameters. Below we briefly summarise the different existing constraints, 
 
 \begin{figure}[t]
 	\centering
 	\includegraphics[height=0.25\textheight,width=0.55\textwidth]{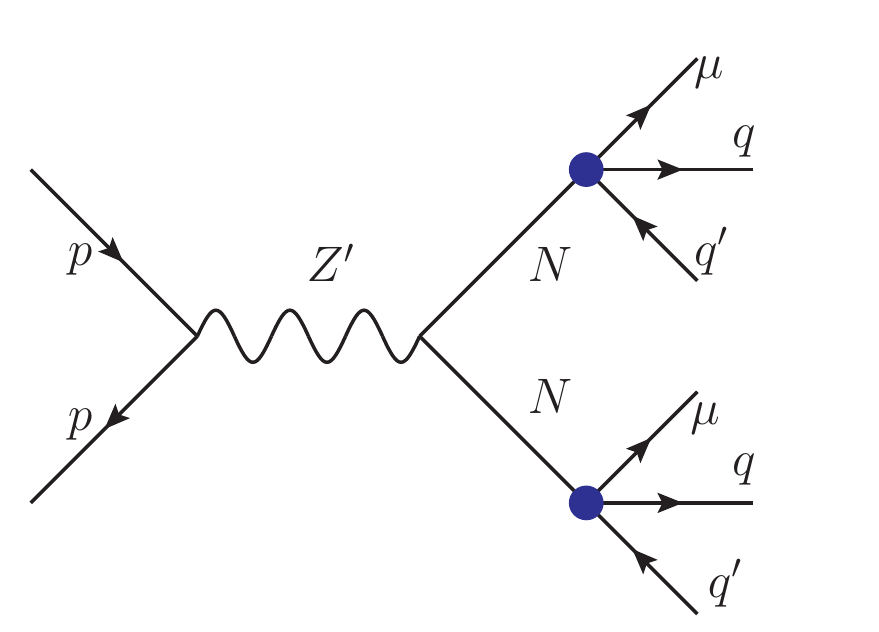}
 	\caption{Feynman diagram representing pair-production of two RHNs via $Z^{\prime}$ mediation and its displaced decays to a $lqq^{\prime}$ final states.\label{fig:feynmandiag}}
 \end{figure}
 \begin{itemize}
 \item {\bf Heavy resonance search:} the search for a massive resonance at LHC decaying to di-lepton/di-jet imposes tight limits on the respective production cross-section. The 13 TeV LHC search for a heavy resonance decaying into two leptons $pp\to Z^\prime \to l^+l^-$  put a constraint $M_{Z^{\prime}} > 5.0$ TeV at 90$\%$ C.L, assuming a 100$\%$ branching ratio of $Z^{\prime}$ decaying into two leptons. For different branching ratios, the limit relaxes. In Fig.~\ref{fig:lhclimit}, we translate the di-lepton constraint from  the LHC on $M_{Z^{\prime}}$, and $g^{\prime}$ plane. We adopt  the following procedure in translating the bound. For the ATLAS  search with $\mathcal{L}=139\,\rm{fb}^{-1}$, we consider ${\sigma^{\text{th}}(pp\to Z^\prime \to l^+l^-)}<\sigma_{obs}$, where the observed limit corresponds to the stringent $95\%$ C.L. limit from \cite{ATLAS:2019erb}. 
{ The CMS limit is set on the relative cross section, $\sigma^{\text{rel}}=\frac{{\sigma \times Br}^{\text{obs}}\big|_{Z^{\prime}}}{{\sigma \times Br}^{\text{obs}}\big|_{Z}}\times 1928\,  \textrm{pb}$. Therefore, to obtain the $95\%$ C.L. limit on ${\sigma \times Br}^{\text{obs}}\big|_{Z^{\prime}}$ we fold $\sigma^{\text{rel}}$ with ${\sigma \times Br}^{\text{obs}}\big|_{Z}/1928$ pb. Here, ${\sigma \times Br}^{\text{obs}}\big|_{Z}$ is the observed cross section in the  $m_{l^+l^-}=60-120$ GeV window, which can be calculated as $N^{\rm{obs}}/(\rm{Acc} \times \rm{Eff} \times\mathcal{L})$\footnote{For the di-electron (di-muon) channel $N^{\rm{obs}}=28194452~(164075) $,  Acc $\times$  Eff$ = 0.176~(0.073)$ and $\mathcal{L}$ =137 (140) $\rm{fb}^{-1}$~\cite{CMS:2021ctt}.}.} In Fig.~\ref{fig:lhclimit}, we represent the $ee+\mu\mu$ combined limit by the purple line from the CMS search and the orange line represents the limit from the ATLAS search. As we consider the results for $\Gamma/M_{Z^{\prime}}=0.6\%$ for the CMS search, we show this by black horizontal dotted line. The dashed green line represents the future sensitivity of HL-LHC with $3000$\, $\rm{fb}^{-1}$~\cite{Ruhr:2016xsg} and the dashed magenta line represents projection for FCC-hh~\cite{Helsens:2019bfw} . We also note that the former  search  at LEP-II \cite{Cacciapaglia:2006pk,SLACE158:2003onx,Electroweak:2003ram,Carena:2004xs} for a massive resonance constrains $Z^{\prime}$ mass and gauge coupling, and thus the $B-L$ breaking scale as  $v_{BL} \equiv M_{Z^\prime}/{(2g^\prime)} \geq 3.45$~TeV. It has been shown by the dashed red line, which is considerably relaxed w.r.t. the present LHC limit.
 
\item
{\bf Search for RHN:}  Other than the constraint on $Z^{\prime}$, the RHN mass and mixings are also constrained, both from neutrino mass measurements as well as direct searches at the LHC. In the present work, we consider relatively low mass RH neutrinos, 10~GeV~$ < M_N < M_W$, which are mainly produced from the $Z^{\prime}$ mediated channel. 
 %\item
 The RHN mass and its mixing with the active neutrinos are tightly constrained from light neutrino mass measurements. As we are working with a  Type-I seesaw scenario with $B-L$ gauge symmetry, the light neutrino mass $m_\nu \simeq {m^2_D}/M_R \simeq V^2_{lN}M_N$ where the active-sterile mixing angle  $V_{lN} \simeq m_D/M_R$. %The sub-eV scale light neutrino mass constraints from $0\nu\beta\beta$ and Tritium beta decay experiments as well as from cosmological observations such as Planck \cite{Ade:2015xua} 
 This fixes the active-sterile mixing, 
\begin{align}
	V_{lN} \approx 10^{-6}\sqrt{\frac{m_\nu / (0.1~\text{eV})}{M_N / (50~\text{GeV})}}.
\end{align}
%
%In any case, we will not restrict ourselves to choose only these values, rather 
{The low mass RHN is further constrained by LHC searches for a heavy neutral lepton \cite{CMS:2022fut}  
via the decay mode $p p \to W \to l N$ and the decay $N \to  l l \nu/ljj$.  The search for a displaced neutral lepton \cite{CMS:2022fut,CMS:2021lzm} in particular constrains active-sterile mixing as $V_{\mu N} < 10^{-2}$ for RHN mass $\mathcal{O}(10)$ GeV. Other searches such as \cite{ATLAS:2018dcj,CMS:2018jxx,CMS:2021dzb} mainly target heavier masses, where the RHN is not displaced, and hence not relevant for our case. }

\item 
{\bf LLP searches from SM Higgs decay:} there are different  exotic decays of the SM Higgs boson which give rise to distinctive signatures. The decay  $H_1\to NN$ followed by the decay of $N$ into $ljj/ll\nu$ final states give rise to displaced lepton/displaced jet signature, as  in the mass range of interest $N$ has a large decay length. The decay is  open kinematically, however it has a very suppressed branching ratio due to our choice of a small $\alpha$. The partial decay width of this decay mode has the form
\begin{equation}
\Gamma(H_1 \to NN)=\frac{3 M_{H_1} {y^M}^2 \sin^2\alpha}{8\pi}  \left(1-\frac{4 M_{N}^2}{M_{H_1}^2}\right)^{3/2},  ~~ \textrm{where}\,~~ y^M=\frac{M_N}{\sqrt{2}v_{BL}}.
\label{eq:htonn}
\end{equation}
 For instance, the branching ratio BR$(H_1\to NN)$ becomes $0.005\%$ for $\sin\theta = 3 \times 10^{-3}$ and $M_N=50$ GeV. There are different CMS and ATLAS searches \cite{CERN-EP-2021-106,CMS:2021uxj,CMS-PAS-EXO-18-003,CMS:2020iwv} to probe exotic decays of the SM Higgs into two LLP states, which can be compared with. This includes a search  for exotic decays of the Higgs boson into LLP in the tracking system~\cite{CERN-EP-2021-106,CMS:2021uxj, CMS-PAS-EXO-18-003,CMS:2020iwv}, or a search for LLP decaying in the ATLAS muon spectrometer~\cite{ATLAS-CONF-2021-032,ATLAS:2019jcm,ATLAS:2018tup}, and in the CMS endcap muon detectors~\cite{CMS:2021juv}. Due to a very suppressed branching ratio,  $M_N$ and  $V_{lN}$  are unconstrained from these searches.
 \end{itemize}
  \begin{figure}[t]
  	\centering
  	\includegraphics[height=0.3\textheight,width=0.55\textwidth]{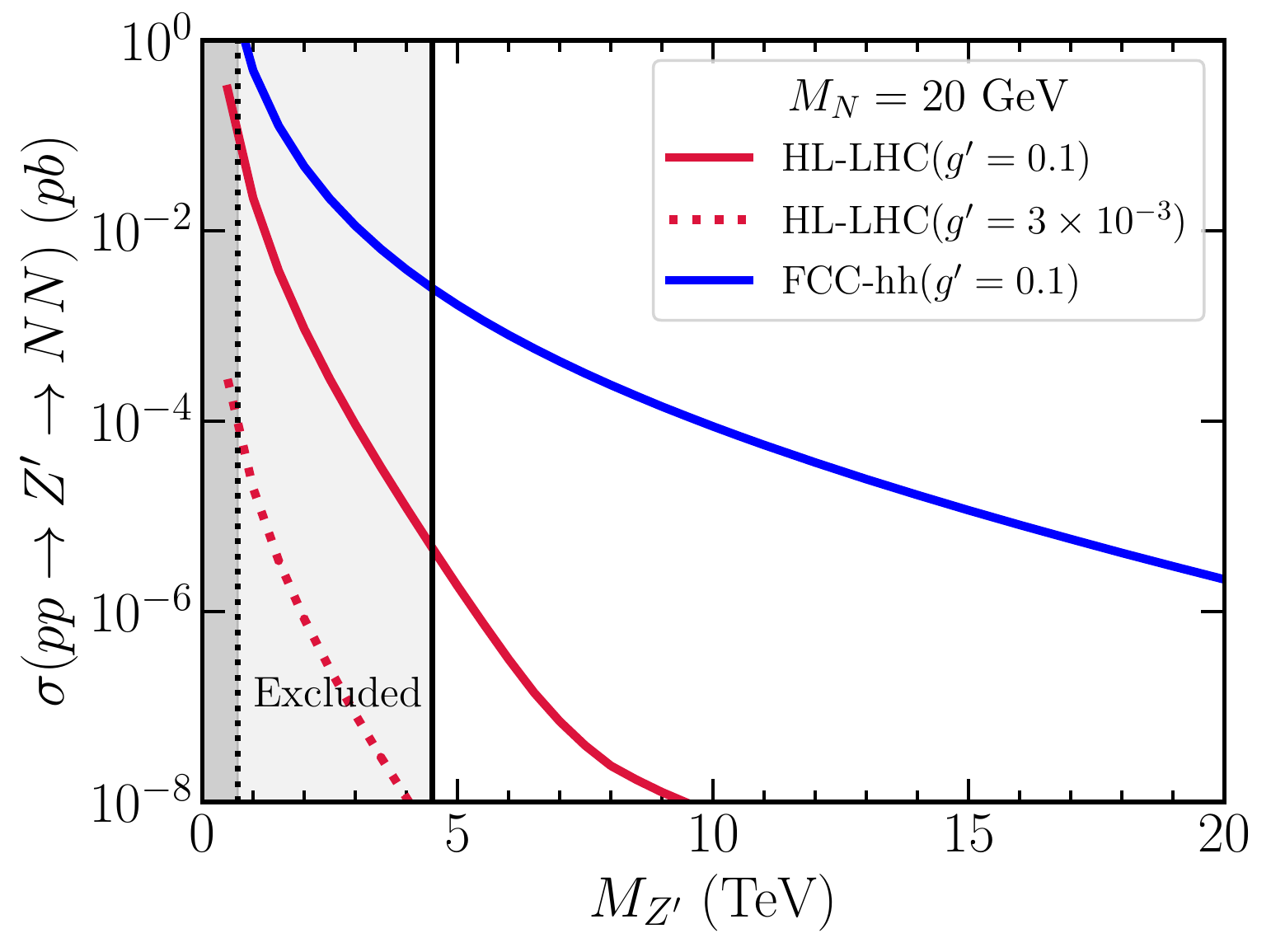}
  	\caption{Cross-section $\sigma (pp\to Z^\prime\to NN)$ as a function of $M_{Z^\prime}$ at the HL-LHC and FCC-hh, for $M_N=20$ GeV. Black solid (dashed) line represents LHC limits for $g^\prime=0.1\ (3\times10^{-3})$ as shown in Fig.~\ref{fig:lhclimit}.}\label{fig:xsvsmzp}
  \end{figure}
\section{Pair production and decay of RHN  at a $pp$ machine \label{sec:HPD}}
 
 The RHN can be pair produced via the processes  $q \bar{q} \to Z^{\prime} \to N N$ and $q \bar{q} \to H_1/H_2 \to N N$, among which the Higgs mediated channels give suppressed cross-sections, as we consider a small SM-BSM Higgs mixing $\sin \alpha$. Therefore, production in pair primarily occurs via $p p \to Z^{\prime} \to N N$ channel. The cross-section depends strongly on the gauge coupling $g^{\prime}$, the mass of $Z^{\prime}$ gauge boson $M_{Z^{\prime}}$, its decay width $\Gamma_{Z^{\prime}}$, as well as the mass of the RHN $M_N$. We show the respective Feynman diagram for this process in Fig.~\ref{fig:feynmandiag}. 
 %---------------------------------------------------------------------------------------------------------------------------------------------------------------------------------------------------------------------------------------------------------------------

 In the left and right panel of Fig.~\ref{fig:xsvsmzp}, we show the pair production cross-section of RHN, $\sigma (pp\to Z^\prime\to NN)$ as a function of $M_{Z^\prime}$ at  the HL-LHC and FCC-hh, respectively. {This cross-section is calculated using MadGraph. For illustration, we assume two different values of couplings, $g^\prime=0.1, 0.003$ and $M_N=20$ GeV. We show the LHC constraints~(as shown in Fig.~\ref{fig:lhclimit}) by the black solid and dotted   lines  for $g^\prime$ values 0.1 and 0.003, respectively. For instance, for the choice of gauge coupling $g^{\prime}=0.003$, the heavy resonance search in $l^+l^-$ decay channel rules out any value of $M_{Z^{\prime}}\lesssim0.7$ TeV. For  $g^{\prime}=0.1$ the exclusion limit reaches higher value, $M_{Z^{\prime}}\gtrsim4.5$ TeV.}  Since in this figure we consider $M_{Z^{\prime}}$ satisfying $M_{Z^{\prime}}>2M_N$, the on-shell production of $Z^{\prime}$ and its decay will dominate the pair-production cross-section $\sigma (p p \to N N)$. As can be seen from the figure,  for the same gauge coupling $g^{\prime}$, the ratio of cross-section $\mathcal{R}=\sigma(p p \to N N)_{\it HL-LHC}/ \sigma(p p \to N N)_{\it  FCC-hh}$ vary over a wide range. For on-shell $Z^{\prime}$ production with narrow width approximation $\Gamma_{Z^{\prime}}/M_{Z^{\prime}}\simeq10\%$, or less, this can be expressed as, 
 \begin{eqnarray}
 \sigma(p p \to N N)_{\it HL-LHC}/ \sigma(p p \to N N)_{\it  FCC-hh} \approx  \sigma(p p \to Z^{\prime})_{\it 14}/ \sigma(p p \to Z^{\prime})_{\it  100} 
 \end{eqnarray}
 With an increase in the c.m. energy from $\sqrt{s}=14$ TeV to 100 TeV, the ratio $\mathcal{R}$ increases by order of magnitude, and the increase is  larger  for higher masses. This occurs as the partonic c.m. energy is larger in a 100 TeV collider 
 compared to the available partonic c.m. energy for 14 TeV.
 %compared to the 14 TeV c.m.energy relevant for HL-LHC, thereby enabling the scope of detection of this signal for heavier $Z^{\prime}$. 

 \begin{figure}
 	%\centering
 	\includegraphics[height=0.3\textheight,width=0.45\textwidth]{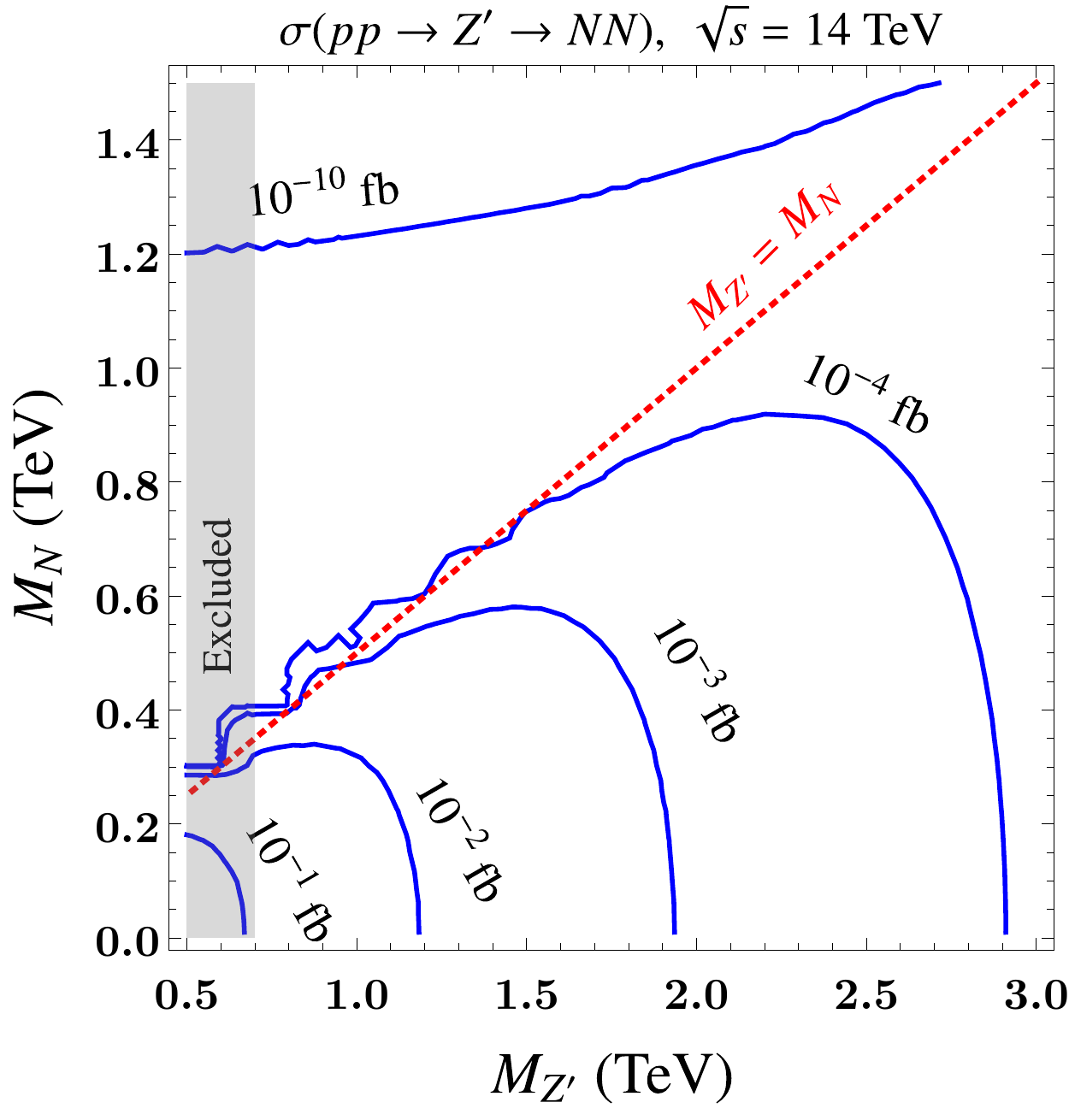}
 	\hspace{0.2cm}
 	\includegraphics[height=0.3\textheight,width=0.45\textwidth]{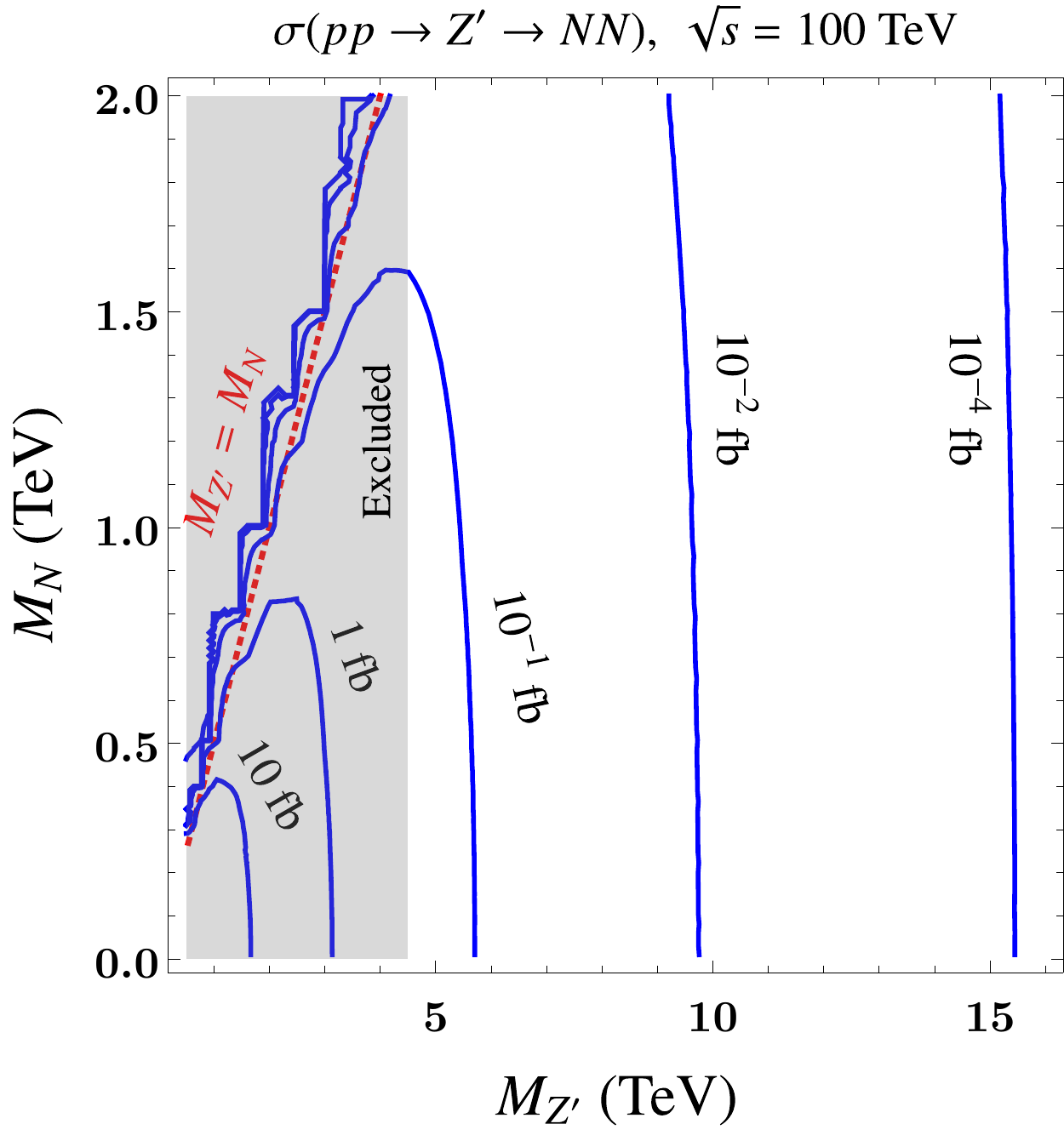}
 	\caption{Cross-section $\sigma (pp\to Z^\prime\to NN)$ as a function of $M_N$ and $M_{Z^\prime}$. Left: For $g^\prime=0.003$  and c.m. energy $\sqrt{s}=14$ TeV. Right: For $g^\prime=0.1$  and c.m. energy $\sqrt{s}=100$ TeV. The vertically shaded region represents the LHC exclusion limit. \label{fig:xsfcc}}
 \end{figure}
 
 We also show the variation of cross-section for HL-LHC and FCC-hh w.r.t. the variation of both $M_N$ and $M_{Z^{\prime}}$  in  Fig.~\ref{fig:xsfcc}. For this figure we consider the same values of gauge coupling that we use for Fig.~\ref{fig:xsvsmzp} The shaded vertical region show the present constraint  from LHC, both in the left and right panel. In the figure of left panel, the cross-section is significantly suppressed for $M_N> M_{Z^{\prime}}/2$, as can be seen from the figure. This happens as $Z^{\prime}$ is not on-shell in this region. For $Z^{\prime}$ with mass $\sim $ TeV and $M_N \ll M_{Z^{\prime}}$, the cross-section can be in the sub-fb range. From the figure of right panel, this can be seen, that for higher $Z^{\prime}$ mass the production cross section is almost  independent of the mass of RHN for the variation of $M_N=100$ GeV to 2 TeV.
% \begin{figure}
 %	\centering
 	%\includegraphics[height=0.25\textheight,width=0.45\textwidth]{mass-z'-cross.pdf}
 %	\caption{Variation of branching ratio of $N$  to different three body decay modes for $M_N <M_W$. }\label{fig:branching}
 %\end{figure}
 
  {\it Decay of RHN:} The RHN interacts with a lepton and a $W$ gauge boson, which is governed by the active-sterile mixing $V_{lN}$. It also interacts with light neutrino and a $Z, H_1$, as well as, $Z^{\prime}$ and BSM Higgs $H_2$.  For the mass range being considered in this work  $M_Z^{\prime}, M_{H_2} \gg  M_{W}, M_Z, M_{H_1} > M_N$, the RHN  $N$  decays pre-dominantly via an off-shell $W,Z,H_1$ states into two SM fermions. The two body decay of RHN into a $(lW, \nu Z, \nu H_{1,2})$ state will only open if mass of $N$ larger than the two body kinematic limit.  The partial decay widths for  $N \to l q q^{\prime}$,  $\nu f \bar{f}$ and $\nu \nu \nu$ decay modes have the following expressions \cite{Bondarenko:2018ptm},
   
  \begin{align}
\Gamma(N \to l_\alpha^-u \bar{d})=N_c |V^{CKM}_{ud}|^2 |V_\alpha|^2\frac{ G_F^2 M_{N}^5}{192 \pi^3} \mathcal{I}(x_{u},x_{d},x_{l})
\label{eq1}
\end{align}
In the above, $\mathcal{I}(x_{u},x_{d},x_{l})= 12 \int_{(x_{d}+x_{l})^2}^{(1-x_{u})^2}\frac{dx}{x} (1+x_{u}^2-x) (x-x_{d}^2-x_{l}^2)\lambda^{\frac{1}{2}}(1,x,x_{u}^2)\lambda^{\frac{1}{2}}(x,x_{l}^2,x_{d}^2)$,  $x_{u/d/l}=\frac{m_{u/d/l}}{M_{N}}$, $\lambda(a,b,c)=a^2+b^2+c^2-2ab-2bc-2ca$,
and  $N_c=3$ is the color factor.
The partial decay width for $N \to l_\alpha^-\nu_\beta l^+_\beta$ with generation index $\alpha \neq \beta$ have the following expression, 
\begin{align}
\Gamma(N \to l_\alpha^-\nu_\beta l^+_\beta)= |V_\alpha|^2\frac{ G_F^2 M_{N}^5}{192 \pi^3} \mathcal{I}(x_{l_\alpha},x_{l_\beta},x_{\nu_\beta})
\label{eq2}
\end{align}
The partial decay width for the other mode with a same $\alpha=\beta$ is given by, 
\begin{align}
\Gamma(N \to \nu_\alpha f \bar{f})= N_c |V_\alpha|^2\frac{ G_F^2 M_{N}^5}{192 \pi^3}
\Big[ C_1^f\Big( (1-14x^2-2x^4-12x^6)\sqrt{1-4x^2}+12 x^4(x^4-1)L(x)\Big) + & \nonumber\\
4C_2^f\Big( x^2(2+10x^2-12x^4)\sqrt{1-4x^2}+6 x^4(1-2x^2+2x^4)L(x)\Big)\Big]
\label{eq3}
\end{align}
Here $x=\frac{m_{f}}{M_{N}}$, $L(x)=\log\Big[ \frac{1-3x^2-(1-x^2)\sqrt{1-4x^2}}{x^2(1+\sqrt{1-4x^2})}\Big]$. The values of $C_1^f$ and $C_2^f$ are given in \cite{Bondarenko:2018ptm}.
 \begin{align}
 \Gamma(N \to \nu \nu \nu)= |V_\alpha|^2\frac{ G_F^2 M_{N}^5}{96 \pi^3} 
 \label{eq3nu}
 \end{align}
 In addition to this there can also be two body decays $N \to l^{\pm} \pi^{\pm}$, however, this is suppressed in our case, as we consider $M_N > 1$ GeV $\sim$ the scale of non-perturbative QCD. Therefore, we do not consider this channel in our analysis. { For $M_N=10-70$ GeV, the dominant decay mode is $N\to\mu jj$ with a branching ratio BR$(N\to\mu jj)\simeq0.5$. Branching ratios for other decay modes are BR$(N\to\nu jj)\simeq0.2$, BR$(N\to\nu_\mu l^+l^-)\simeq0.10$, BR$(N\to\mu e \nu_e+ \mu \tau \nu_\tau)\simeq0.15$ and BR$(N\to \nu \nu \nu)\simeq0.05$. }
    %We show the respective branching ratios for these few different channels in Fig.~\ref{fig:branching}. \MMb{MM: discussion about the branching of $N$ to be added.} 
%%%%%%%%%%%%%%%%%%%%%%%%%%%%%%%%%%%%%%%%%%%%%%%%%%%%%%%%%%%%%%%%%%%%%%%%%%%%%%%%
\section{RHN decay in inner tracker and  muon spectrometer \label{indmuspec}}
%%%%%%%%%%%%%%%%%%%%%%%%%%%%%%%%%%%%%%%%%%%%%%%%%%%%%%%%%%%%%%%%%%%%%%%%%%%%%%%%%
We are interested in a signature which contains a pair of RHNs produced from $Z^{\prime}$ decay and the decay of $N$ to a  $N \to l q q^{\prime}$ final state. To be specific, we consider the $N \to \mu q q^{\prime}$ final state and analyse the model signature. {Signatures with an electron in the final state $N \to e q q^{\prime}$ can also potentially be explored by similar search strategies.} Specifically,  we consider two scenarios: (a) when the $N$ decays within the inner detector (ID) of the {HL-LHC/FCC-hh}, and (b) in the muon spectrometer (MS). {The search strategies for these two different decay modes differ widely}.  We will also consider a combination of these two, when one RHN decays in the ID and other in the MS.  In order to analyse the  {model signatures}, we simulate the events using the following steps. {We use the FeynRules~\cite{Alloul:2013bka,Christensen:2008py} model file and Universal FeynRules Output (UFO)~\cite{Degrande:2011ua} corresponding to Ref.~\cite{Deppisch:2018eth}, which we use  in combination with the Monte Carlo event generator {\tt MadGraph5aMC$@$NLO} -v2.6.7~\cite{Alwall:2014hca} to generate events at the parton level}.  The FeynRules~\cite{Alloul:2013bka,Christensen:2008py} model file  and UFO is publicly available from the FeynRules Model Database at~\cite{FeynrulesDatabase}.  For every signal sample, we generate {50000} signal events with {\tt MadGraph5aMC$@$NLO} -v2.6.7, where we use {NN23LO1} PDF set \cite{Buckley:2014ana}. We set max jet flavour at 5 to take account of the $b$ quark contribution in the PDF.  We then pass the generated parton level events on to PYTHIA v8.235~\cite{Sjostrand:2014zea} which handles the initial and final state { radiation, parton showering, hadronization, and heavy hadron decays}. The clustering of the events and { jet formation}  are performed by FastJet v3.2.1~\cite{Cacciari:2011ma}. We consider a Cambridge-Aachen jet algorithm \cite{Dokshitzer:1997in}  for jet clustering with radius parameter $\mathcal{R}=1.0$. {We analyse events at the generator level with PYTHIA v8.235.}  %\textcolor{red}{FIXME: check code versions, describe RIVET etc.}
%%%%%%%%%%%%%%%%%%%%%%%%%%%%%%%%%%%%%%%%%%%%%%%%%%%%%%%%%%%%%%%%%%%%%%%%%%
\subsection{Decay Probability of $N$ \label{decayprob}}
\begin{figure}[t]
\centering
	\includegraphics[height=0.3\textheight,width=0.6\textwidth]{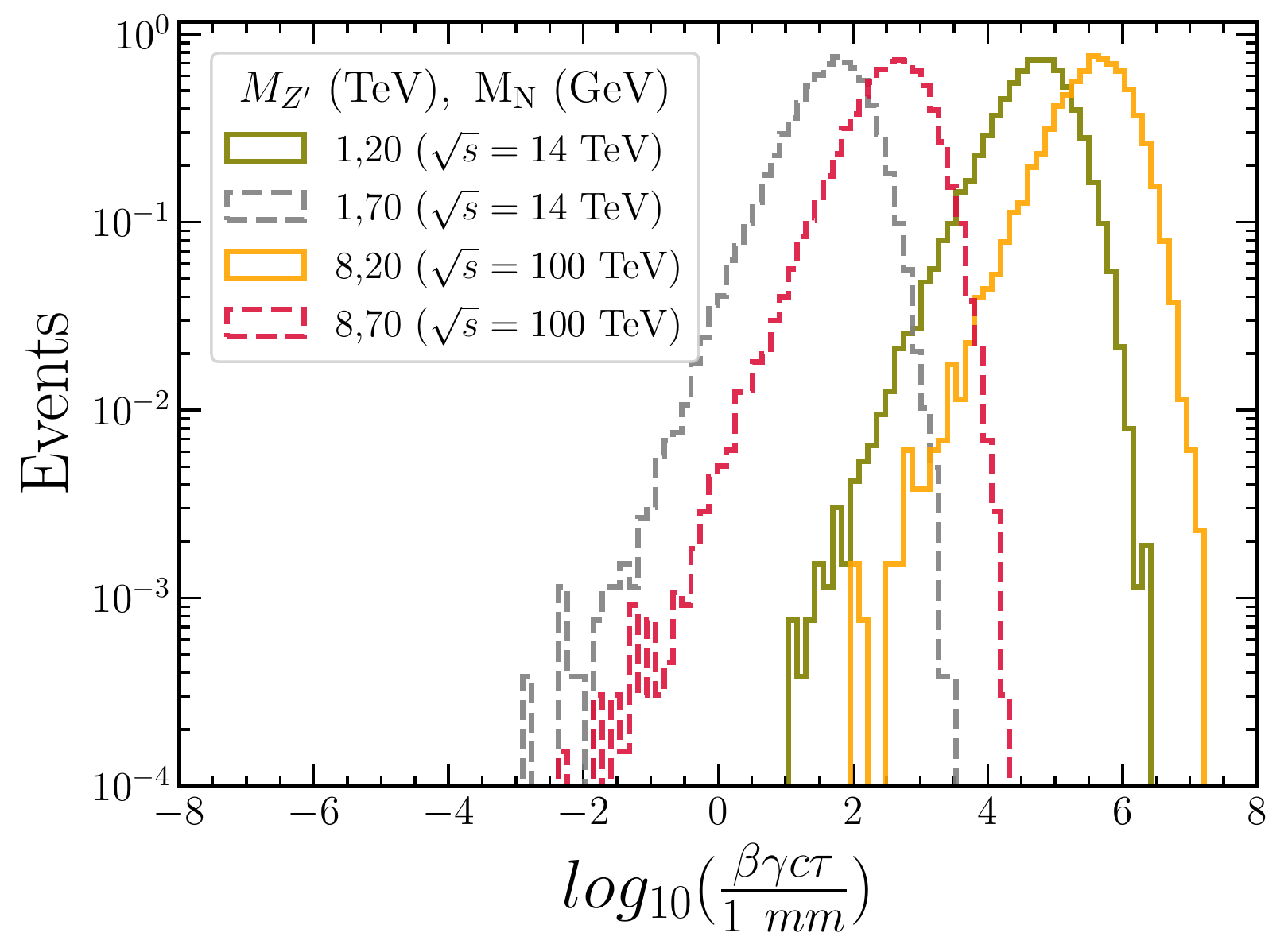} 
	\caption{{Distribution of the decay length of $N$ at the HL-LHC and at FCC-hh for the considered masses of $Z^{\prime}$ and  $N$}.}
	\label{fig:tauN}
\end{figure}
For $M_N<M_W$ and for the viable  range of active-sterile mixing that satisfies light neutrino mass constraint,  the RHN $N$ undergoes displaced decays with a displacement ${\it l} \sim mm$ or even longer. 
$N$ decays via off-shell $W,Z,H$ states with  the decay length \footnote{The contribution from the Higgs in partial decay width is an order of magnitude  smaller and hence we do not consider this.} %{\MMb{\bf{MM: include the expression}}}},
\begin{equation}
\tau_N\simeq 25 \times \Big( \frac{10^{-6}}{|V_{\mu N}|}\Big)^2 \times \Big( \frac{100 ~\rm{GeV}}{M_N}\Big)^5~\rm{mm}\label{eq:taun1}
\end{equation}
%\begin{equation}
%\tau=1/\Gamma^{total}_N, ~~~ \Gamma^{total}_N=\Gamma_N(N \to l q q^{\prime}) +\Gamma_N(N %\to \nu l^+l^-)+ \Gamma_N(N \to \nu l^+_{\alpha} l^-_{\beta}) \label{eq:taun}
%\end{equation} %\end{document}
%where the analytic expressions of the partial decay widths $\Gamma_N(N \to l q q^{\prime}), \Gamma_N(N \to \nu l^+l^-)$ and $\Gamma_N(N \to \nu l^+_{\alpha} l^-_{\beta})$ have been given in Eq.~\ref{eq1}, Eq.~\ref{eq2} and Eq.~\ref{eq3}, respectively.  
In Fig.~\ref{fig:tauN}, we show the distribution of the decay length of $N$ in the lab frame for $\sqrt{s}=14$ TeV and $\sqrt{s}=100$ TeV for few benchmark values of $M_{Z^{\prime}}$, and $M_N$, which we later on use in the analysis.
 {For this distribution we generate  $p p \to Z^{\prime} \to N N$ in MadGraph, and to keep the  information of decay length,  use the  decay-in-flight option of  MadGraph.}
  As can be seen, in the majority of events, decay length of $N$ is more than a $mm$, thereby giving rise to a decay vertex considerably displaced from the production vertex of $N$. 

As we have two RHNs from the channel $p p \to Z^{\prime} \to N N$, therefore the two RHNs can decay in different parts of the HL-LHC and FCC-hh detectors. The probability that both the RHNs decay within a distance interval $(L_1, L_2)$ from their production vertex is given by
\begin{eqnarray}
\mathcal{P}(L_1, L_2, \sqrt{s}, M_N,M_{Z^\prime},\theta)=\int db_1 db_2 \ f(\sqrt{s}, M_N,M_{Z^\prime},b_1,b_2) \ \prod_{i=1}^{2}\Big[e^{\frac{-L_1}{b_ic\tau(\theta)}}-e^{\frac{-L_2}{b_ic\tau(\theta)}}\Big]
\label{prob1}
\end{eqnarray} 	
where $c\tau$ is proper decay length, $\theta$ represents the active-sterile mixing  and $b_{1,2}=(\beta\gamma)_{1,2}$ are the boost factors of the two RHNs. 
%\footnote{{$b_1$ and $b_2$ are the boost factors of the two $N$s.}}
  and $f(\sqrt{s}, M_N,M_{Z^\prime},b_1,b_2)$ is the probability distribution function  of the boost.  
%\item
The probability that one RHN decay within interval $(L_1, L_2)$ and the other within interval $(L_3, L_4)$ is given by
\begin{eqnarray}
\mathcal{P}(L_1, L_2,L_3,L_4,\sqrt{s}, M_N,M_{Z^\prime}, \theta)= \ &2 \times\int db_1 db_2 \ f(\sqrt{s}, M_N,M_{Z^\prime},b_1,b_2) \ [e^{\frac{-L_1}{b_1c\tau(\theta)}}-e^{\frac{-L_2}{b_1c\tau(\theta)}}] \times\nonumber \\& [e^{\frac{-L_3}{b_2c\tau(\theta)}}-e^{\frac{-L_4}{b_2c\tau(\theta)}}] \label{prob2}
\end{eqnarray} 	
%\end{itemize}
The decay probability of one RHN decaying within length $L_1-L_2$ then becomes, 
\begin{eqnarray}
\mathcal{P}(L_1, L_2,\sqrt{s}, M_N,M_{Z^\prime},\theta)= \ &\int db  \ f(\sqrt{s}, M_N,M_{Z^\prime},b) \ [e^{\frac{-L_1}{b c\tau(\theta)}}-e^{\frac{-L_2}{b c\tau(\theta)}}]  \label{prob3}
\end{eqnarray} 	
where $b$ is the boost factor of the respective RHN. We use the boost distributions of  $N$ from  the HepMC file \cite{Buckley:2019xhk}, which has been generated by showering the LHE events~\cite{Alwall:2006yp} from MadGraph.
\begin{table}
	\centering
	\begin{tabular}{|lll|}
		\hline
		\multicolumn{3}{|l|}{\hspace{3cm}Detector Geometry }                            \\ \hline
		\multicolumn{1}{|l|}{}   & \multicolumn{1}{l|}{HL-LHC} & FCC-hh \\ \hline
		\multicolumn{1}{|l|}{Inner detector (ID)} & \multicolumn{1}{l|}{(2-300) mm}   &  (25-1550) mm   \\ \hline
		\multicolumn{1}{|l|}{Calorimeter (CAL)} & \multicolumn{1}{l|}{(2000-4000) mm}    &    (2700-4700) mm \\ \hline
		\multicolumn{1}{|l|}{Muon Spectrometer (MS)} & \multicolumn{1}{l|}{(4000-7000) mm}    &  (6000-9000) mm   \\ \hline
	\end{tabular}
	\caption{Coverage of geometric acceptance of various sub-detectors of the HL-LHC and FCC-hh.}
	\label{tab:detectorgeometry}
\end{table}
\begin{figure}
	\includegraphics[height=0.25\textheight,width=0.5\textwidth]{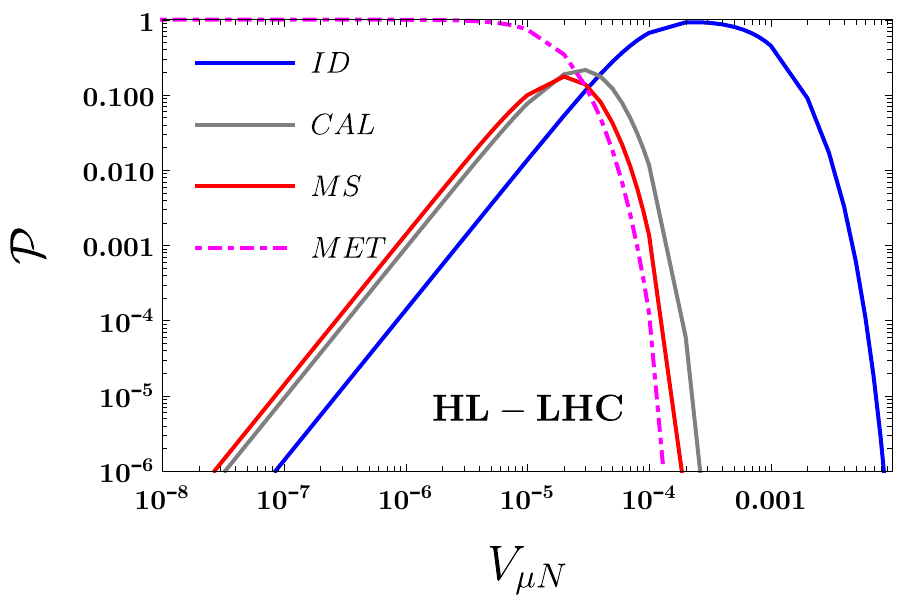}
	\includegraphics[height=0.25\textheight,width=0.5\textwidth]{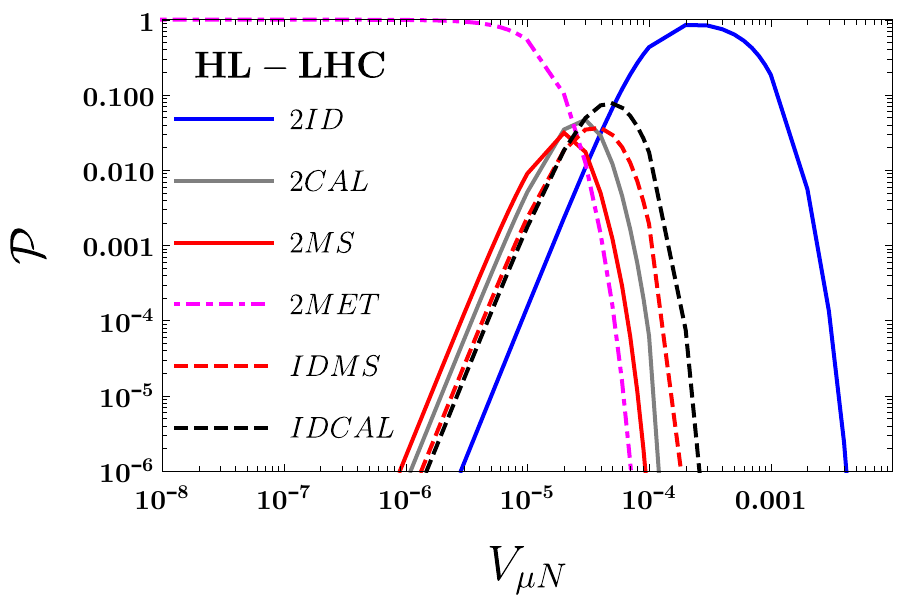}
	\caption{Decay probability of {one RHN} (left) and {two RHN}~(right) in different parts of the HL-LHC detector. For this figure, we consider $M_N=20$ GeV, $M_{Z^\prime}=1$ TeV. ID, MS, CAL and MET represent the decay of $N$ in the inner detector, muon spectrometer, calorimeter and outside of the detector, respectively. In the right panel, IDMS~(IDCAL) represents the joint probability of one $N$ decaying  in the ID and the other decaying in the MS~(CAL). \label{fig:prob} }
\end{figure}

{To estimate the probability of decay of $N$ in various parts of the HL-LHC and FCC-hh detectors, we adopt a simplistic approach. In Table.~\ref{tab:detectorgeometry}, we present the fiducial volume of the sub-detectors, where the decay of RHN leads to observable signatures. For Inner detector~(ID) we demand RHN decays  in between 2 mm and 300 mm. In this region, the vertex reconstruction efficiency is large~\cite{ATLAS:2019jcm}.
 %which corresponds to the region with good ID-vertex reconstruction efficiency~\cite{ATLAS:2019jcm}. 
  The decays of RHN in ID leads to multi-track displaced vertex. Similar to ID the decay of $N$ in the MS creates a number of tracks. For this we consider that RHN decays within 4000-7000 mm where muon ROI trigger efficiency is higher~\cite{ATLAS:2018tup}. Hadronic decays of RHN in the outer edge of the ECal or in the HCal leads to a distinct signatures marked by significantly large energy deposit in HCal compared to that in the ECal~\cite{ATLAS:2019qrr}. We however do not consider such signature in the present analysis, rather restrict to RHN decay in ID and in MS. For the FCC-hh detector coverage of various sub-detectors used in the analysis, we refer the readers  FCC-CDR~\cite{FCC:2018vvp}. }

\begin{figure}[h]
		\includegraphics[height=0.25\textheight,width=0.5\textwidth]{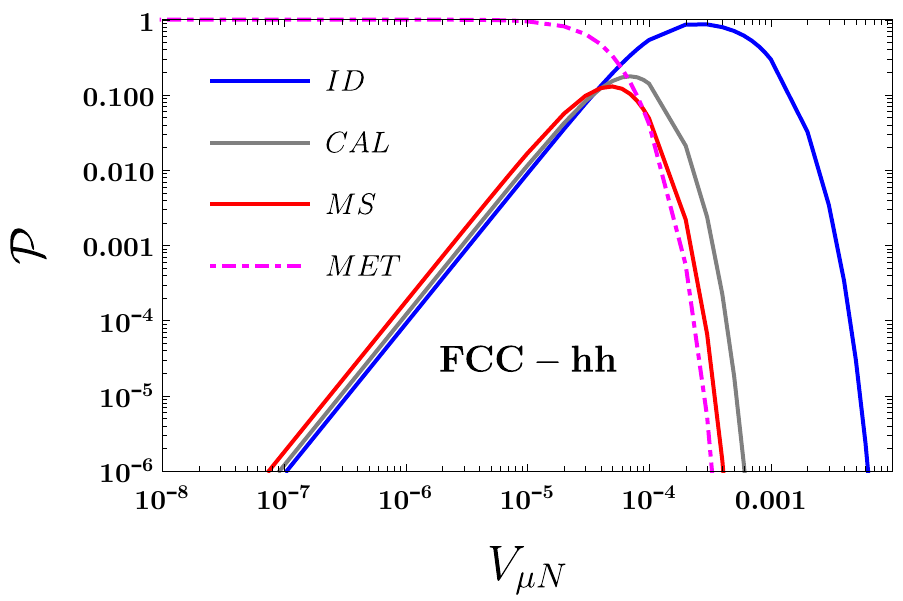}
	\includegraphics[height=0.25\textheight,width=0.5\textwidth]{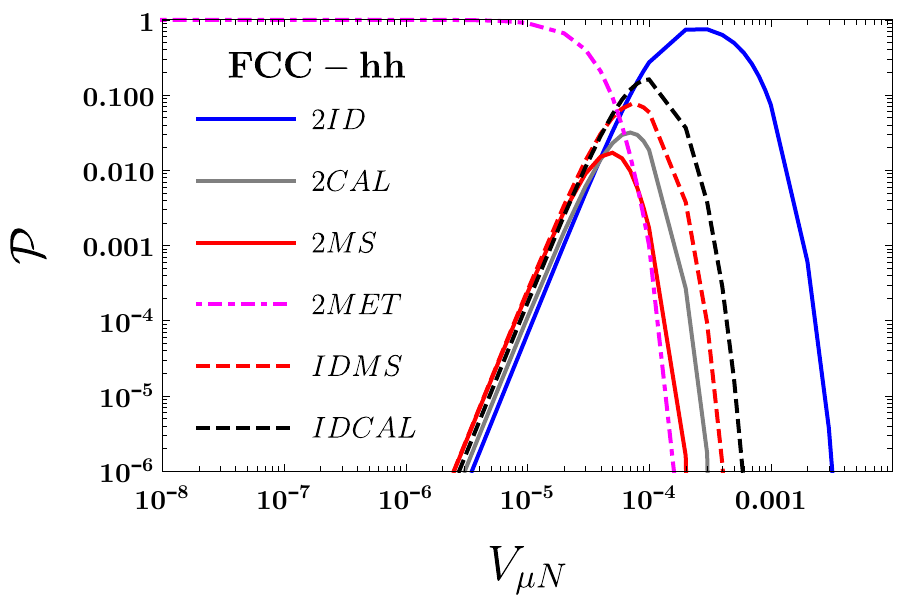}
\caption{As Fig.~\ref{fig:prob} but shows the decay probability for the FCC-hh detector with  $M_N=20$ GeV and $M_{Z^\prime}=8$ TeV.\label{fig:probfcc}}
\end{figure}

{In Fig.~\ref{fig:prob}, we show the probability of the RHNs decaying in different modules of the HL-LHC detector. The left panel represents the decay probability of a single $N$ in the detector and in the right panel, we show the probability for both the $N$s to undergo displaced decays. The acronyms  ID, CAL, MS, MET refer to  the decay of RHN inside the inner detector, the calorimeter, the muon spectrometer, and outside of the CMS and ATLAS detector, respectively. From the figure, it is evident that the probability is maximal if the RHN decays inside the inner detector or outside of the detector, and it is minimal, if the decay happens in the muon spectrometer. The probability for other configurations, such as if the decay occurs in the calorimeter, or among the two $N$s one decays in the inner detector and another in calorimeter/muon spectrometer is somewhat in between. We show the respective probabilities for the FCC-hh in Fig.~\ref{fig:probfcc}.    }

{Fig.~\ref{fig:prob} and Fig.~\ref{fig:probfcc} illustrate useful information about the respective geometric probabilities  of the RHN to undergo displaced decays. A naive comparison between the results derived for HL-LHC and FCC-hh shows that while the maximum decay probability of one/two $N$  decaying  within ID is very similar $\simeq 95\%$, the probability for one $N$ decaying in the ID and the other in MS is somewhat larger in FCC-hh than HL-LHC. A large probability along with a higher c.m. energy and a higher achievable integrated luminosity will result in a significantly  large  number of events,  that can be observed at the FCC-hh.  In the subsequent sections, we present a detailed discussion of the observable signal events at HL-LHC and FCC-hh, while  taking into account both the kinematic and geometric cut efficiencies.}

%%%%%%%%%%%%%%%%%%%%%%%%%%%%%%%%%%%%%%%%%%%%%%%%%%%%%%%%%%%%%%%%%%%%%%%%%%%%%%%%%%%%%%%%%%%
\subsection{Signal description for $N$ decaying in the ID \label{sigdesid}}
\begin{figure}[t]
	\centering
	\includegraphics[height=0.3\textheight,width=0.5\textwidth]{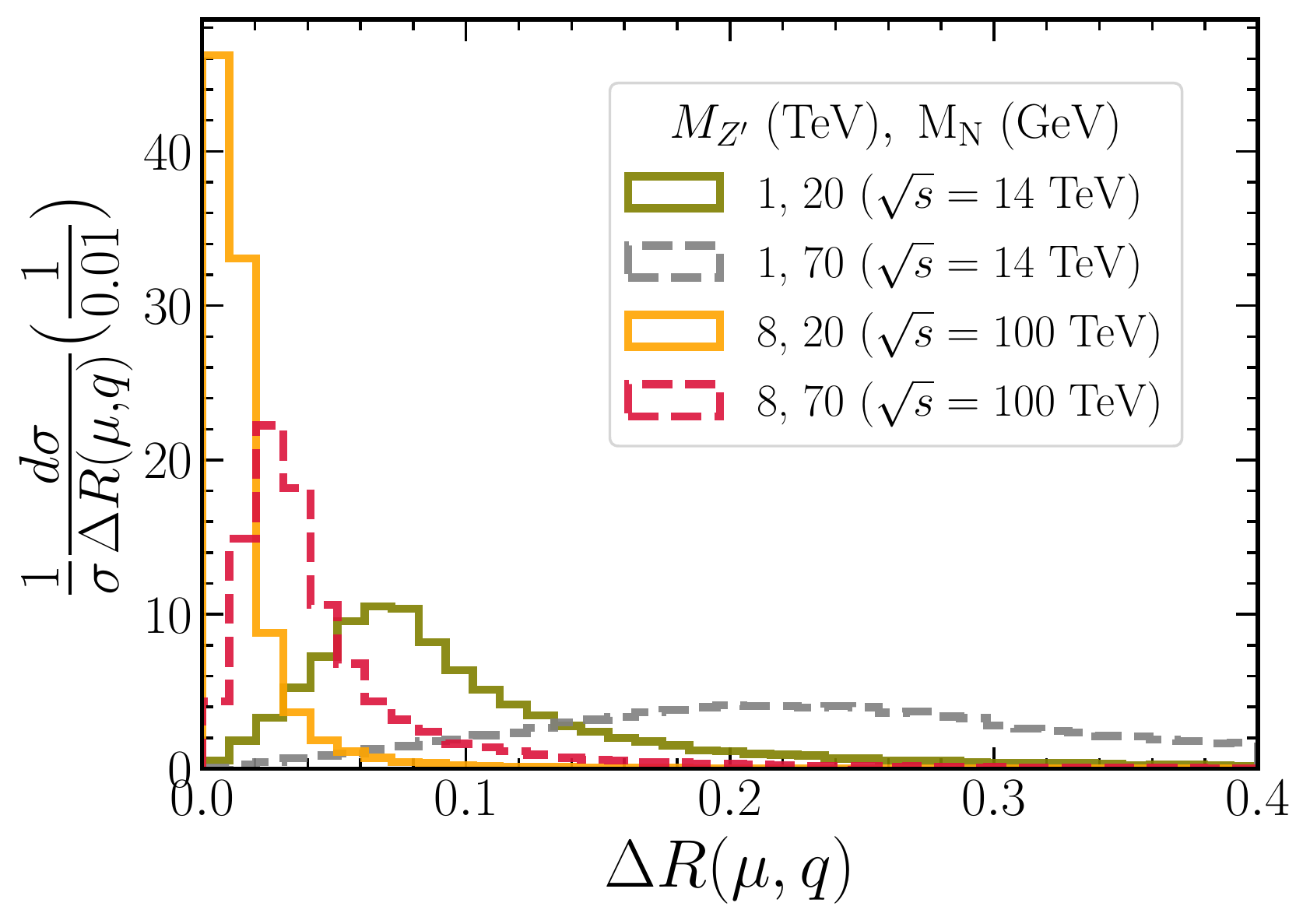}
	\caption{$\Delta R$ separation between the $\mu$ and  closest $q$ at the partonic level. 	}
	\label{fig:drmuq}
\end{figure}

\begin{figure}[t]
\centering
	\includegraphics[height=0.25\textheight,width=0.45\textwidth]{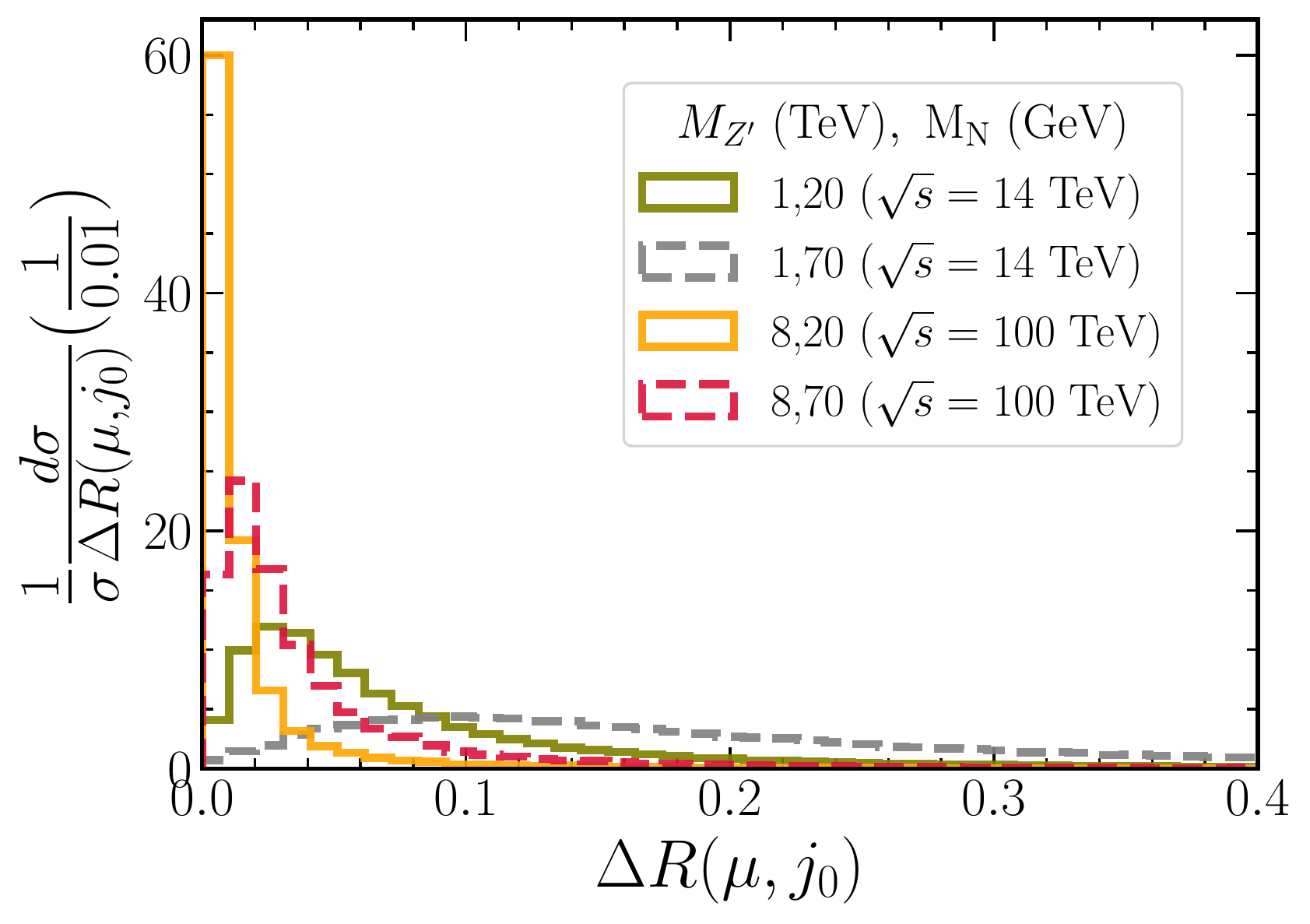}
	\includegraphics[height=0.25\textheight,width=0.45\textwidth]{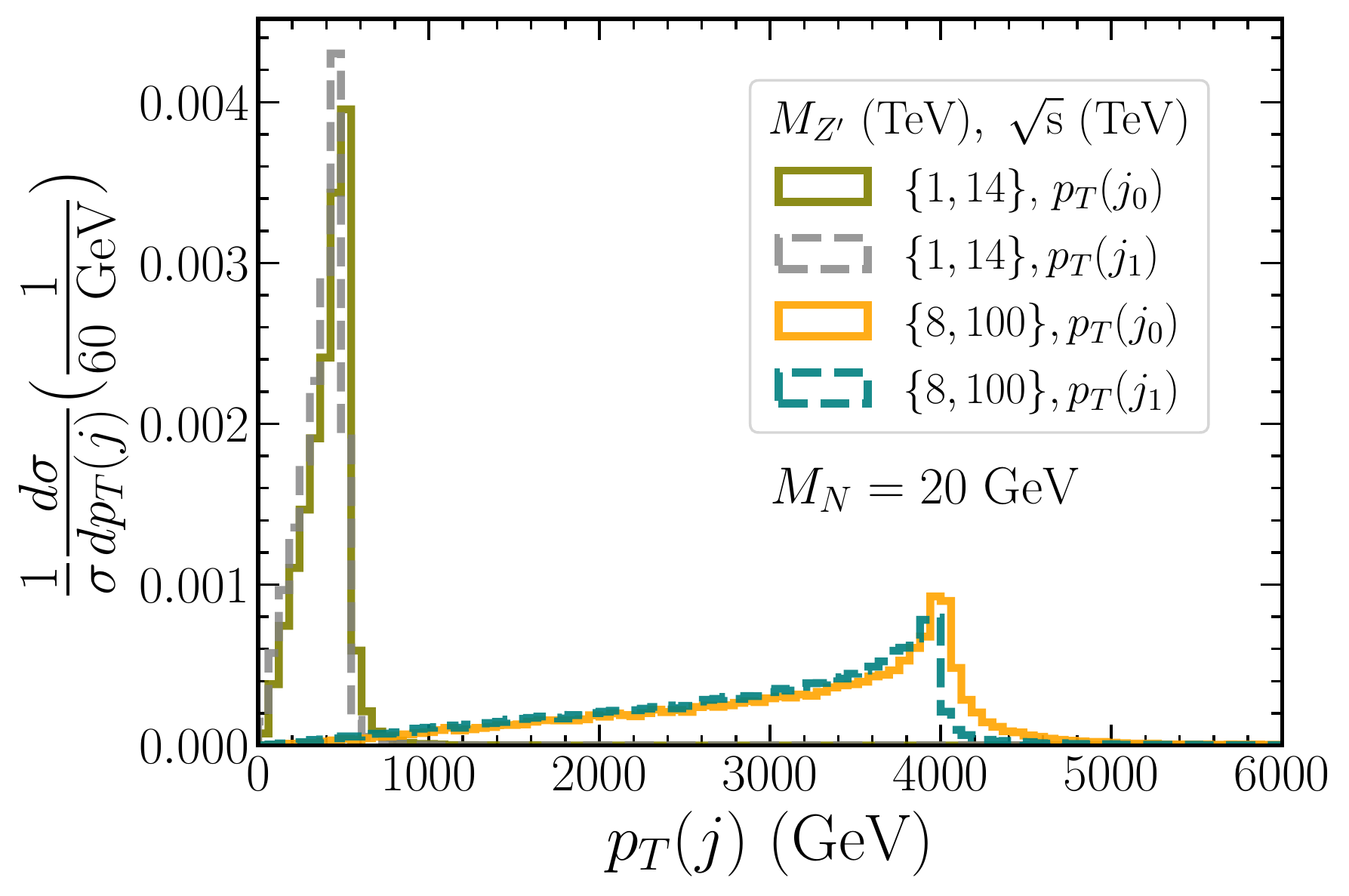}
	\caption{Left: $\Delta R$ separation between the $\mu$ and  leading jet. Right: Transverse momentum distribution of the leading and sub-leading fat-jet. 	}
	\label{fig:deltaRmuj0}
\end{figure}

 %\MMb{MM: the following few line of texts  need a bit of re-arrangement.} 
 {As we are focussing on a heavier $Z^{\prime}$ with mass $M_{Z^{\prime}}\ge 1$ TeV,  and lighter $N$ with masses $10-70$ GeV satisfying $M_N <M_W$, the $N$ hence will be boosted, resulting in collimated decay products\footnote{$M_N<10$ GeV is not motivated due to more SM background for displaced vertex signatures.}. In Fig.~\ref{fig:drmuq}, we show the $\Delta R=\sqrt{(\Delta \eta)^2+(\Delta\phi)^2}$ separation between the muon and the closest quark at the parton level, originating from the decay of a RHN. The $\Delta R$ separation between two closest quarks also shows similar features. As can be seen, in most of the events the muon-quark separation $\Delta R (\mu,q)\ll 0.4$, where  $\Delta R=0.4$  has often been utilised in experimental analyses as the isolation criterion between final states.} Instead of abiding by the  selection criterion with jet radius $\mathcal{R}=0.4$ and  standard isolation criterion $\Delta R(\ell, j), \Delta R(j,j) >0.4$, we hence demand a large jet radius $\mathcal{R}=1.0$,  referred to as {\it fat-jet} \cite{CMS:2017bcq}. Since we consider a large jet-radius and do not impose any specific isolation criterion, in the majority of the events,  the lepton hence will be part of the {\it fat-jet}. Extending further, we also show the distributions of $\Delta R$ after jet clustering. {In left panel of Fig.~\ref{fig:deltaRmuj0}, we show the $\Delta R$ separation of the closest  $\mu$ from the leading jet, for both $\sqrt{s}=14$ TeV and $\sqrt{s}=100$ TeV.}  It is evident from the figure that the $\Delta R$ separation between the $\mu$ and the $j$ is $\Delta R(\mu,j) \ll 0.4$, {which mimics the partonic distribution, shown in Fig.~\ref{fig:drmuq}}. %\MMr{The distribution at the parton level i.e., separation between the lepton and closest quark, as well as quark-quark separation also display similar characteristic feature, see Fig.~\ref{f}.} 
% Therefore imposing the standard isolation criterion between the  lepton and jet as $\Delta R > 0.4$ will drastically reduce the signal selection efficiency.  In  majority of the events, the lepton will be inside the fat-jet .  
 
 As we have pointed out before, for the considered   RHN state with mass $M_N<M_W$,  the eV scale light neutrino mass demands a very suppressed active-sterile mixing $V_{lN} \sim 10^{-6}$, thereby leading to a macroscopic decay length for $N$. Therefore, the final state originating from $p p \to Z^{\prime} \to N N$ decay would be 
 \begin{itemize}
 \item
$p p \to Z^{\prime} \to N N \to \underbrace{\mu j j }~~\underbrace{\mu j j} \to J^{\it dis}_{fat}~~J^{\it dis}_{fat} $ 
\label{eq:disfatN}
\end{itemize}
In the above $J^{\it dis}_{fat}$ represents the {\it fat-jet} originating from $N$ decay,  which is also  displaced. Since the {\it fat-jet} is formed by the  decay products of $N$, which is considerably lighter than $Z^{\prime}$, satisfying $\frac{M_N}{M_{Z^{\prime}}}\ll1$, the $N$ and hence the leading and sub-leading jet have a high transverse momentum. We show the $p_T$ distribution of leading and sub-leading jets in {the right panel of Fig.~\ref{fig:deltaRmuj0}. As can be seen, for $M_{Z^{\prime}}=1$ TeV and $M_N=20$ GeV, the distribution peaks around $p_T \sim M_{Z^{\prime}}/2=500$ GeV. For $M_{Z^{\prime}}=8$ TeV, the  peaks appear  at a higher value, $p_T \sim M_{Z^{\prime}}/2=4$ TeV.} This high transverse momentum of leading and sub-leading jets will be useful in designing the selection cuts. 

We also note that the proposed signal can also be extended to include $ejj$ final state. This however depends strongly on the active-sterile mixing $V_{eN}$. With  $V_{eN}\ll V_{\mu N}$, the contribution from $ejj$ channel will be small. Here, we adopt a simplistic approach and consider only $V_{\mu N}\neq 0$, therefore do not consider $ejj$ state in {\it fat-jet} description.

%%%%%%%%%%%%%%%%%%%%%%%%%%%%%%%%%%%%%%%%%%%%%%%%%%%%%%%%%%%%%%%
\subsection{Signal description for $N$ decaying in the MS \label{sigdesms}}
If the RHN decays within the muon spectrometer, then one can not use the information on the number of tracks in the ID, $p_T$ of track, and other variables such as deposited energy in the calorimeter. Instead of performing  an analysis based on observables of a {\it fat-jet}, we rather present a track based analysis.  The final state originating from $p p \to Z^{\prime} \to N N $ would be considered as
\begin{itemize}
\item
$p p \to Z^{\prime} \to N N \to \underbrace{\mu q q^{\prime} }~~\underbrace{\mu q q^{\prime}} \to \mu^{\it dis} \mu^{\it dis} + X^{\it dis} \to \underbrace{ track_1+track_2+...track_n}+ \underbrace{Y}$
\label{sigdescrims}
\end{itemize}
where $\mu^{\it dis}$ represents a  muon, produced from the RHN displaced decays. In the above, $X$ corresponds to all other final state particles, including hadrons, electrons, photons etc. which are generated due to RHN decays as well as due to showering.  We represent all the charged particles generated from the decay of RHN  as $track_1, track_2,..track_n$ and $Y$ represent any neutral particle in the final state that does not leave any footprint of a charged track in the muon spectrometer.  

For these above mentioned model signatures, we first perform the analysis for HL-LHC with c.m. energy $\sqrt{s}=14$ TeV considering an integrated luminosity  $\mathcal{L}=3\, \rm{ab}^{-1}$ and later investigate the discovery  prospect for FCC-hh with $\sqrt{s}=100$ TeV with  $\mathcal{L}=30\, \rm{ab}^{-1}$. To  evaluate the number of  events for the above mentioned two signal descriptions, we implement  both the geometric and kinematic cuts, the details  of which are provided in Section.~\ref{hllhcproj}, and Section.~\ref{fcchhproj}. The geometric cuts take into account the probability of RHN to decay within a specified  region of HL-LHC/FCC-hh detector. For the geometric cut efficiency, we follow  the prescription described  in Section.~\ref{decayprob}, and use   Eq.~(\ref{prob1}), Eq.~(\ref{prob2}), and Eq.~(\ref{prob3}). As kinematic cuts, we use  following few variables -  $p_T$ of jet, $p_T$ of associated tracks, $\eta$ of jet and tracks. Below, we briefly outline the adopted procedure, which we implement for the analysis:
\begin{itemize}
\item %\underline{Results from simulation:}\\
We implement both kinematic and geometric cuts to evaluate the number of signal events that can be detected. Let $g$ and $k$ denote the  geometric and kinematic cuts. We denote  the corresponding probabilities for an event to pass $g$ and $k$ by  $\epsilon_g$ and $\epsilon_k$, respectively, and  $\epsilon_{g\&k}$ is the {probability} that an event pass both $g$ and $k$. 
 The {probability} that an event known to pass the kinematic cuts $k$, will also pass $g$ is given as (using conditional probability),
\begin{equation} \epsilon_{g|k}=\epsilon_{g\&k}/\epsilon_{k}. \label{condiprob}\end{equation}
% \pagebreak
%\item \underline{ Results using analytic formula:}\\
To evaluate the above, we use  the boost distribution of RHN, undergoing displaced decays.  In Eq.~(\ref{prob1}), Eq.~(\ref{prob2}), and Eq.~(\ref{prob3}), we consider that  $f(\sqrt{s}, M_N,M_{Z^\prime},b_1,b_2)$ is the boost distribution of $N$s from  those events that satisfy the kinematic cut. %\MMb{To calculate this, we pass the LHE file from MadGraph  into Pythia v8 for showering and hadronization, where we implement the kinematic cuts. 
We evaluate the geometric cut-efficiency using the above-mentioned equations and   the kinematic cut-efficiencies has been calculated  in PYTHIA v8.235~\cite{Sjostrand:2014zea}.  %Finally  $f(\sqrt{s}, M_N,M_{Z^\prime},b_1,b_2)$ }. 
With this, we obtain
\begin{equation}
 \epsilon_{g|k}={\mathcal{P}}(L_1, L_2, \sqrt{s}, M_N,M_{Z^\prime},\theta).
 \label{eqnp}
 \end{equation}
Using Eq.~(\ref{condiprob}), the final cut-efficiency including both the geometric and kinematic cuts becomes
  \begin{equation}\epsilon_{g\&k} = \epsilon_{g|k} \times \epsilon_{k}={\mathcal{P}}(L_1, L_2, \sqrt{s}, M_N,M_{Z^\prime},\theta)\times \epsilon_{k}, \label{final}\end{equation}
  where the probability  ${\mathcal{P}}$ in the above has been simulated from  the event samples that satisfy the kinematic cuts. 
\end{itemize}
The cross-sections after cut for the two signal descriptions are evaluated as
\begin{equation}
\sigma( p p \to N N \to J^{\it dis}_{\it fat} J^{\it dis}_{\it fat}) \big |_{after-cut}=\sigma_p \times \mathcal{P}(L_1, L_2, \sqrt{s}, M_N,M_{Z^\prime},\theta)\times \epsilon_{k}, \\
\label{finalcrossid}
\end{equation}
and 
\begin{equation}
\sigma( p p \to N N \to \underbrace{ track_1+..track_n}+ \underbrace{Y}) \big |_{after-cut}=\sigma_p \times \mathcal{{P}}(L_1, L_2, \sqrt{s}, M_N,M_{Z^\prime},\theta)\times \epsilon_{k},\\
\label{finalcrossms}
\end{equation}
where $\mathcal{P}(L_1, L_2, \sqrt{s}, M_N,M_{Z^\prime},\theta)$ and $\epsilon_{k}$ represent the corresponding geometric and kinematic cut-efficiencies for RHN decay, respectively. In the above, $\sigma_p$ is the partonic cross-section for $p p \to Z^{\prime} \to N N \to \mu j j \mu j j$. %and 
  We have explicitly checked that the cut-efficiencies following this procedure match with the cut-efficiencies obtained with a full Pythiav8.235 based numerical simulation with a mismatch  $ < \mathcal{O}(5\%)$. 

%%%%%%%%%%%%%%%%%%%%%%%%%%%%%%%%%%%%%%%%%%%%%%%%%%%%%%%
\section{Projection for HL-LHC \label{hllhcproj}}
%%%%%%%%%%%%%%%%%%%%%%%%%%%%%%%%%%%%%%%%%%%%%%%%%%%%%%%%%
To evaluate the discovery prospect of  $N$, which  decays in the inner detector/muon spectrometer of the HL-LHC detector, we use  kinematic variables - transverse momentum $p_T$ and pseudo-rapidity $|\eta|$ of jets, and $p_T$, $|\eta|$ of tracks. We  first obtain  the results  demanding  displaced decays of two $N$s, which give distinctive signatures with low background. This is referred as 2IDvx or 2MSvx events, depending on whether  $N$ decay in the ID or in MS. We also show the projection  relaxing  the tight requirement of exact two displaced vertices, and analyse the signature with at least one $N$ undergoing displaced decay (referred as 1IDvx, and 1MSvx events). Note that all the results are  calculated under the zero background assumption. For the HL-LHC analysis we consider $M_{Z^\prime}=1$ TeV, $g^\prime=0.003$. 
\begin{figure}[t]
	\centering
	\includegraphics[height=0.25\textheight,width=0.45\textwidth]{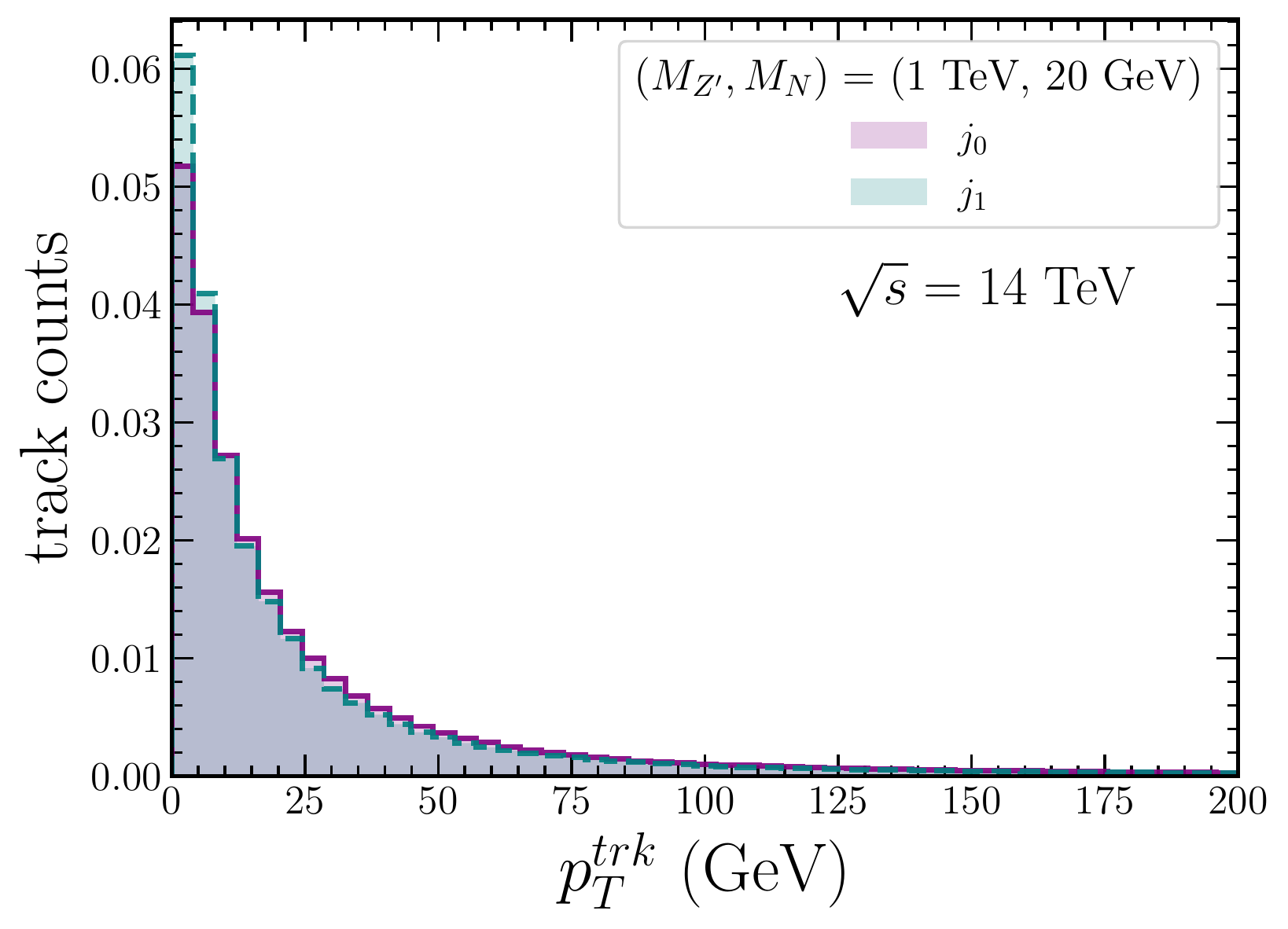}
	\includegraphics[height=0.25\textheight,width=0.45\textwidth]{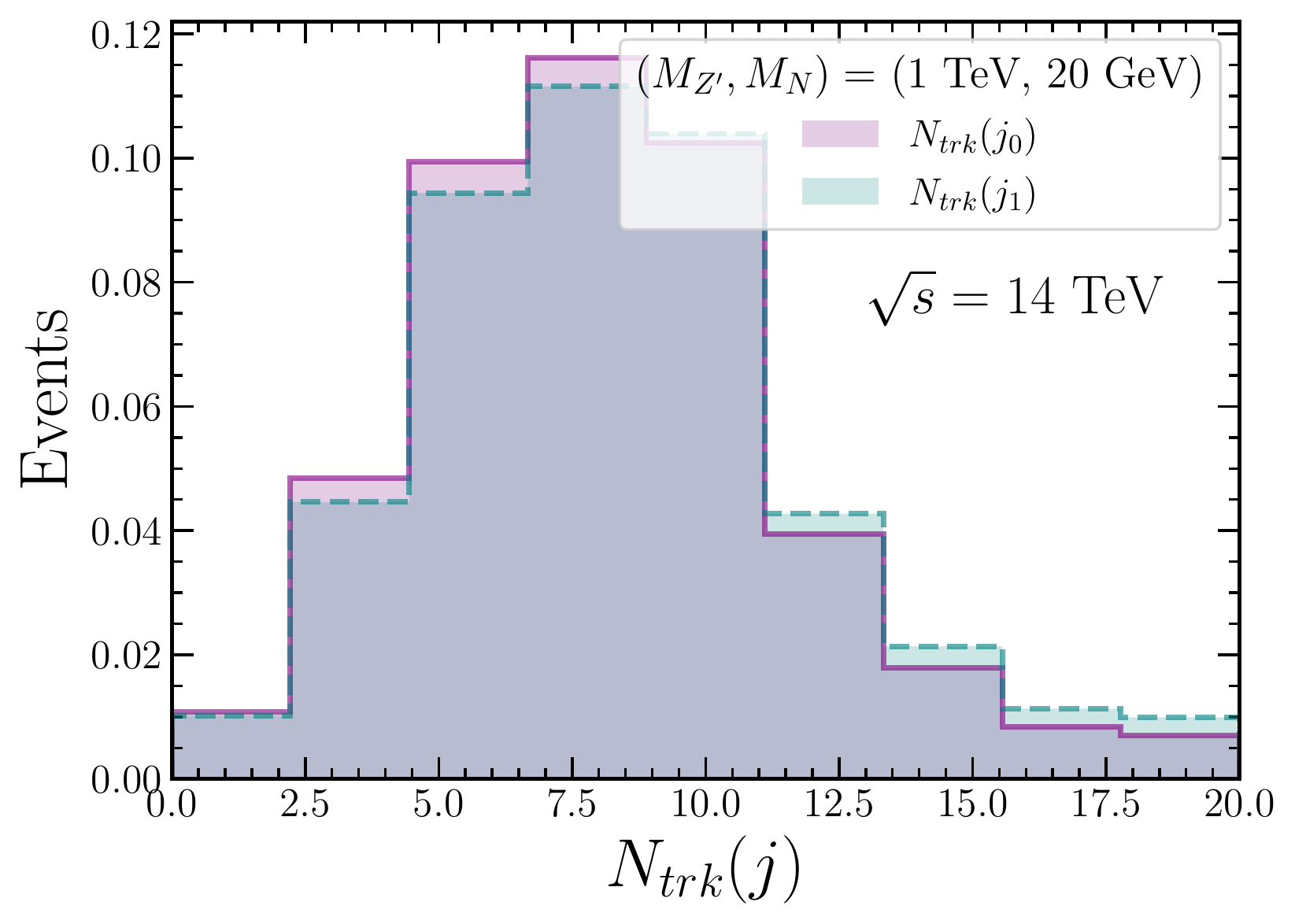}
	\caption{Left panel: $p_T$ of tracks associated to leading and sub-leading jets. Right panel: number of tracks associated to $j_{0,1}$ for HL-LHC. }
	\label{fig:ntrack_hllhc}
\end{figure}
\begin{figure}[h]
	\includegraphics[height=0.30\textheight,width=0.45\textwidth]{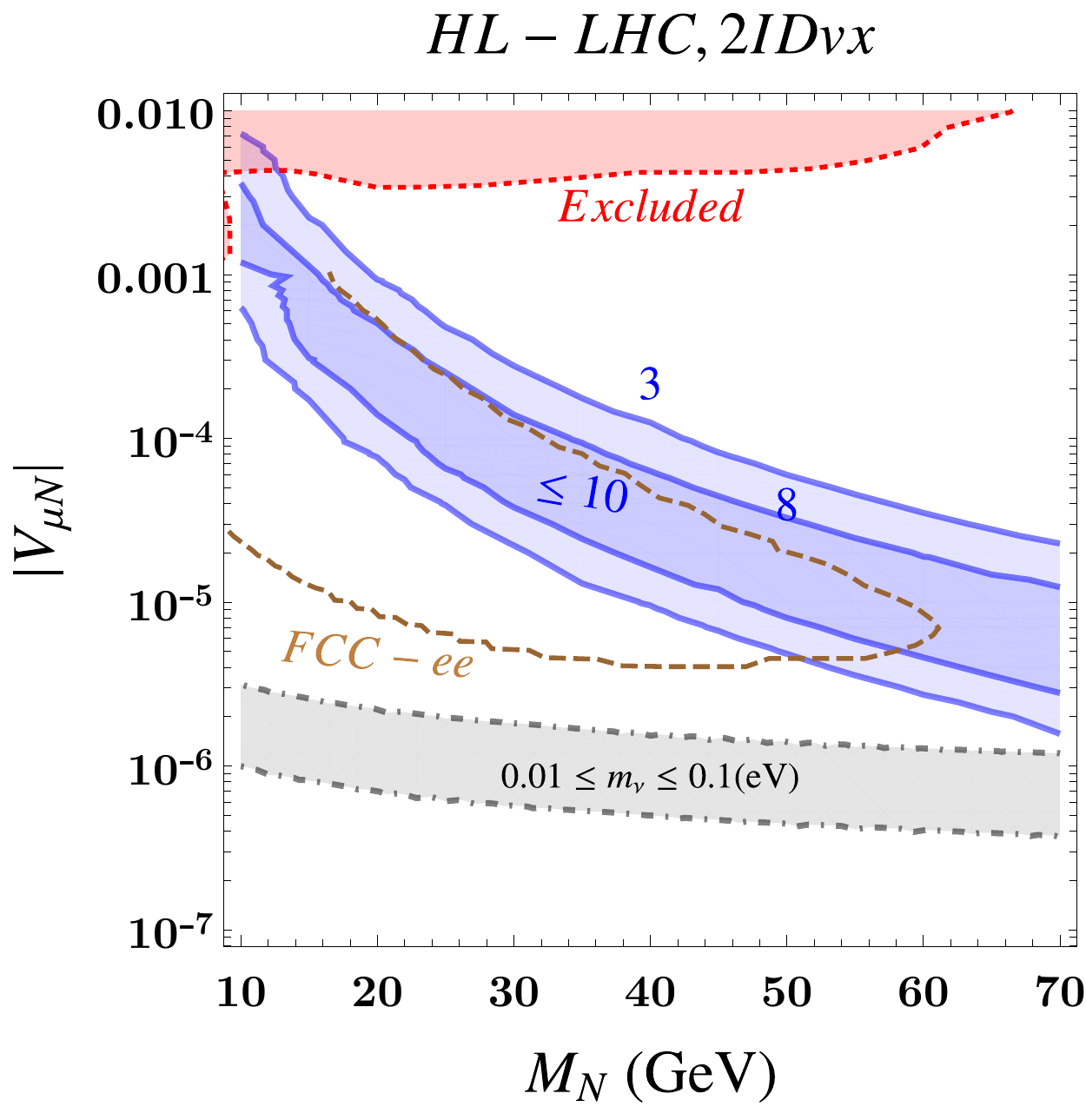} 
	\includegraphics[height=0.30\textheight,width=0.45\textwidth]{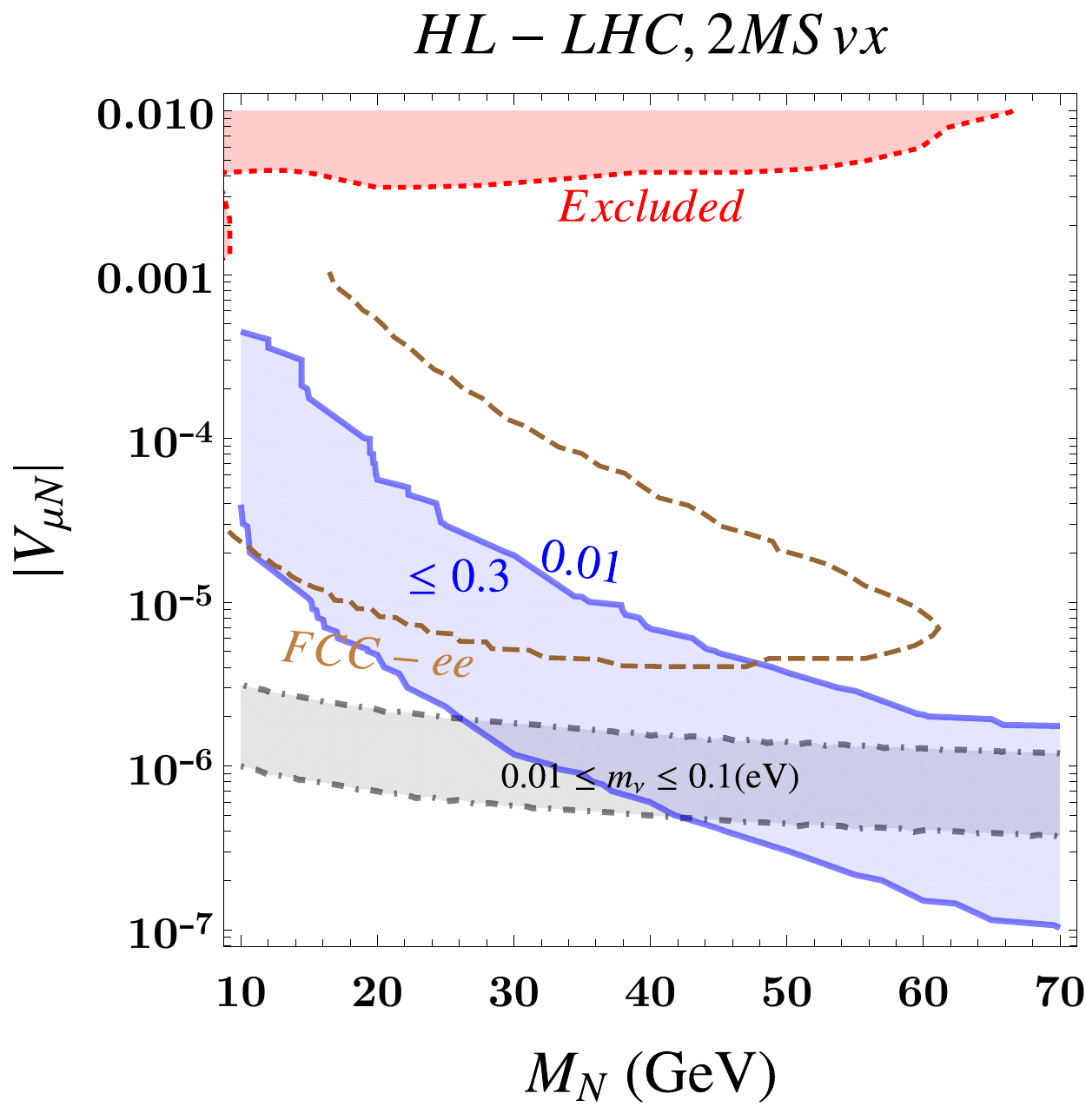} 
	\caption{Contours  of  2IDvx (left) and 2MSvx~(right) events that can be achieved  with $\mathcal{L}=3\,\rm{ab}^{-1}$. A  maximum of $N_{events}=10$ events can be observed in 2IDvx event category. The brown dashed line represents the FCC-ee projection \cite{Blondel:2014bra,Bolton:2019pcu}, and the red shaded region represents the excluded region from \cite{ATLAS:2019kpx,CMS:2018iaf}.
	\label{fig:eventsforhllhc2dis}}
	\end{figure}
	
	\begin{figure}[h]
	\includegraphics[height=0.30\textheight,width=0.453\textwidth]{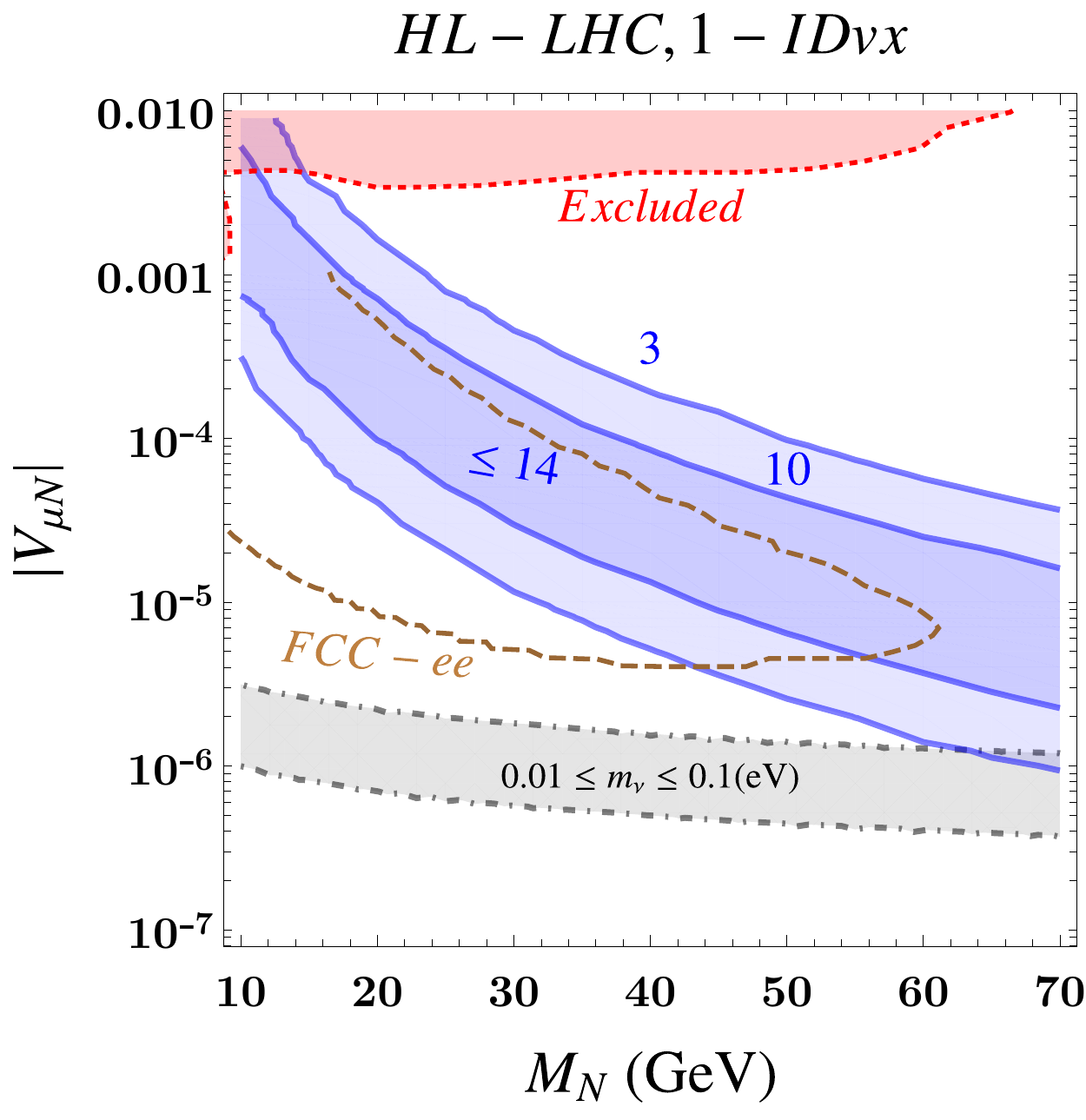} 
	\includegraphics[height=0.30\textheight,width=0.453\textwidth]{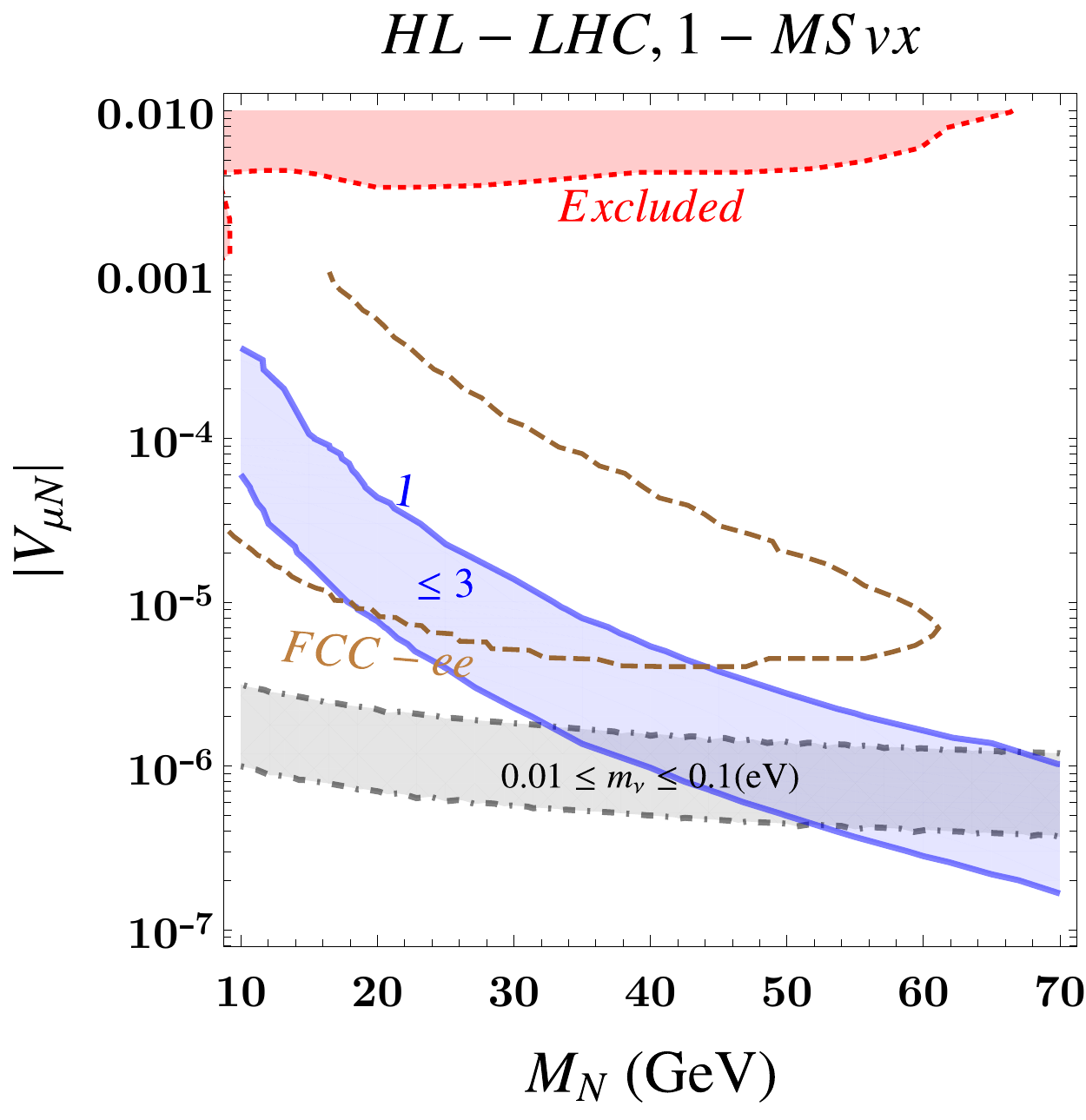} 
	\caption{Contours of  1IDvx ~(left) and 1MSvx (right) events with $\mathcal{L}=3\,\rm{ab}^{-1}$ at the HL-LHC. 
	\label{fig:eventsforhllhc1dis}}
	\end{figure}
%%%%%%%%%%%%%%%%%%%%%%%%%%%%%%%%%%%%%%%%%%%%%%%%%%%%%%%%%%%%%%%%%%%%%%%%%%%
\subsection{Decay vertex in the inner-detector (IDvx): \label{idvx}}
We consider the following sets of cuts for the analysis of two displaced vertices in the ID,
\begin{itemize}
\item
{\it C1:} Both the RHN decay  within $ID_{L_1}=2$ mm and $ID_{L_2}=300$ mm. %We also present the analysis considering displaced decays of at least one RHN
\item
{\it C2:}  The leading and sub-leading fat-jet  $j_{0,1}$ has to satisfy  $|\eta(j)| <4.5$ and $p_T(j)\ge150$ GeV. The strong cut on jet-$p_T$ is motivated by the ATLAS analysis of boosted RHN \cite{ATLAS:2019isd} and other fat-jet analyses~\cite{Banerjee:2018yxy,Dolan:2012rv}.
\item
{\it C3:}  Additionally for each of the leading and sub-leading fat-jet, number of associated track with $p_T\ge1 \text{GeV}$, $|\eta|<2.5$  should be  $n_{trk} \ge4$.  This can be easily satisfied, as majority of events have a $p_T(track)>1$ GeV, as is evident from the left panel of Fig.~\ref{fig:ntrack_hllhc}. The number of tracks associated to the leading and sub-leading jet peaks around $\sim 7$, hence the selection criterion on number of tracks is easily satisfied. These criteria are inspired from the ATLAS search  for long-lived particles decaying into displaced hadronic jets in ID~\cite{ATLAS:2019jcm}.
%(Here $n_{trk}$ is number of tracks associated with the jet.) 
\item
{\it C4:}  Finally we select the events if  number of jets $n_j\ge2$ and the above-mentioned cuts are satisfied for both  the leading and sub-leading {\it fat-jets}. 
%ass IDvx cut.)
\end{itemize}
As we outline in Section.~\ref{decayprob},  we consider the boost distribution of the two $N$s from the events that  satisfy the above mentioned kinematic cuts {\it C2-C4} and evaluate the geometric cut-efficiency as $ \mathcal{P}(L_1, L_2, \sqrt{s}, M_N,M_{Z^\prime},\theta)$ using  Eq.~\ref{prob1}, and Eq.~\ref{prob2}. In the left panel of  Fig.~\ref{fig:eventsforhllhc2dis}, we show the {contours for $N_{events}=3$ and $ 8$ events}  in $|V_{\mu N}|$ and $M_N$ plane by blue solid lines. A  maximum number of $10$   2IDvx events can be observed with the luminosity $\mathcal{L}=3\, \rm{ab}^{-1}$. {In the same plot, we also show the projection for  FCC-ee in the channel $e^+e^-\to Z\to\nu N$ by the brown dashed line, derived in \cite{Bolton:2019pcu,Blondel:2014bra}. The region shaded in red  is  excluded from CMS and ATLAS searches in both prompt and displaced leptonic decay signatures of RHNs~\cite{Bolton:2019pcu,ATLAS:2019kpx,CMS:2018iaf}.} We also show the active-sterile neutrino mixing required for  light neutrino mass $m_{\nu}$ in between $0.01-0.1$  eV.  %\MMb{MM: citations etc to be included...}

We also estimate the number of events demanding  decay of at least one $N$ in the ID \footnote{The other $N$ can decay in any region of the detector, and we are not tagging its decay.} and show the result in the left panel of Fig.~\ref{fig:eventsforhllhc1dis}. For this, we consider only the probability of one RHN decay within $2-300$ mm using Eq.~\ref{prob3} and the kinematic cut-efficiencies as $\epsilon_k \sim 100\%$. We checked explicitly that more than $90\%$ events satisfy the above mentioned kinematic cuts. With the requirement of at least one $N$ decaying in the ID, the probability is slightly larger, which leads to maximum number of  events as ${\it N_{events}=14}$, that can be obtained with $3\, \rm{ab}^{-1}$ of data.  
%%%%%%%%%%%%%%%%%%%%%%%%%%%%%%%%%%%%%%%%%%%%%%%%%%%%%%%%%%%%%%%%%%%%%%%%%%%
\subsection{Decay vertex in the muon spectrometer (MSvx): \label{msvx}} 
We consider that $N$ has a long lifetime and it decays to  $\mu q q^{\prime}$ in the muon-spectrometer. We consider the following sets of selection cuts:
\begin{itemize}
\item {\it C1:} We demand both  the RHN decay  occur within the muon-spectrometer between the length $MS_{L_1}=4000$ mm, and $MS_{L_2}=7000$ mm. {The chosen length interval corresponds to the outer edge of the HCal and the middle station of muon
chambers where the muon RoI trigger efficiency is higher~\cite{ATLAS:2018tup}.}
\item {\it C2:} For each of the tracks originating from RHN decay vertex we impose $p_T(track)>1$ GeV, $ |\eta(track)| <2.7$ and $n_{trk} \ge4$.
\item {\it C3:} Finally we select the events if two  decay vertex and the associated tracks from the vertex satisfy the above selection criterion. {We also demand $\Sigma_{{track}} P_T(track)>60$ GeV, which along with the cut $n_{trk} \ge4$  ensures a significant number of hits in the muon-spectrometer.} 
\end{itemize}
{We evaluate the geometric cut-efficiencies following Eq.~\ref{prob1} and Eq.~\ref{prob2} using  the event samples that satisfy selection cuts {\it C2-C3}. The kinematic and geometric cuts have been designed based on the ATLAS searche~\cite{ATLAS:2018tup}.} In the right panel of Fig.~\ref{fig:eventsforhllhc2dis}, we show the number of events for the 2MSvx events. We also show the number of events demanding decay of at least one $N$ in the muon spectrometer in the right panel of Fig.~\ref{fig:eventsforhllhc1dis}, where we consider only the geometric cut-efficiency. The notable difference between the 2IDvx events and 2MSvx events is that the number of events in the later case reduces significantly. This can be understood by referring Fig.~\ref{fig:prob}, where we show the probability of two $N$ decaying  in different region of the HL-LHC detector. As can be seen that the probability of two $N$ decaying in the muon spectrometer is much smaller $\mathcal{P} \sim 1\%$, compared to the probability of two $N$ decaying in the inner-detector, where probability is $\mathcal{P} >  99\%$. This order of magnitude difference in the geometric cut-efficiency is primarily responsible \footnote{{The kinematic cut-efficiency is more than 90$\%$ for both 2IDvx and 2MSvx events}.}  for the  reduction in  number of 2MSvx signal events. The prospect  of detection of  1MSvx events is higher in HL-LHC compared to 2MSvx events, as can be seen by comparing right panel of Fig.~\ref{fig:eventsforhllhc2dis} and Fig.~\ref{fig:eventsforhllhc1dis}. {A maximum of $N_{events}=3$ events can be obtained with 3 $\rm{ab}^{-1}$ of data.}

We also show the events where one of two $N$s decays in the inner-detector and the other $N$ decays in the muon spectrometer, which we refer as {MSID} event category. The geometric probability for this is somewhat similar to  the respective probability of 2MSvx events, as can be seen from Fig.~\ref{fig:prob}. We show the event contours  in Fig.~\ref{fig:eventsforidms} {assuming $100\%$ kinematic cut-efficiency.} 
%\MMb{MM: a line or two about trigger for MSvx}
%%%%%%%%%%%%%%%%%%%%%%%%%%%%%%%%%%%%%%%%%%%%%%%%%%%%%%%%%%
\section{Projection for FCC-hh \label{fcchhproj}}
%%%%%%%%%%%%%%%%%%%%%%%%%%%%%%%%%%%%%%%%%%%%%%%%%%%%%%%%%%%%%%%%%%%%%%%%%%%
We consider the same decay channel and signal description as we consider for HL-LHC and present results for $M_{Z^\prime}=8 \ \rm{TeV}, \ g^\prime=0.1$. The necessary details about the  geometry of the inner detector, muon spectrometer that we consider for the analysis  has been given in Table.~\ref{tab:detectorgeometry}. {We implement similar set of kenematic cuts as used for HL-LHC study in Sec.\ref{hllhcproj}. Here we use harder $p_T$ cuts for fat-jet.}
\begin{figure}[t]
	\centering
	\includegraphics[height=0.25\textheight,width=0.45\textwidth]{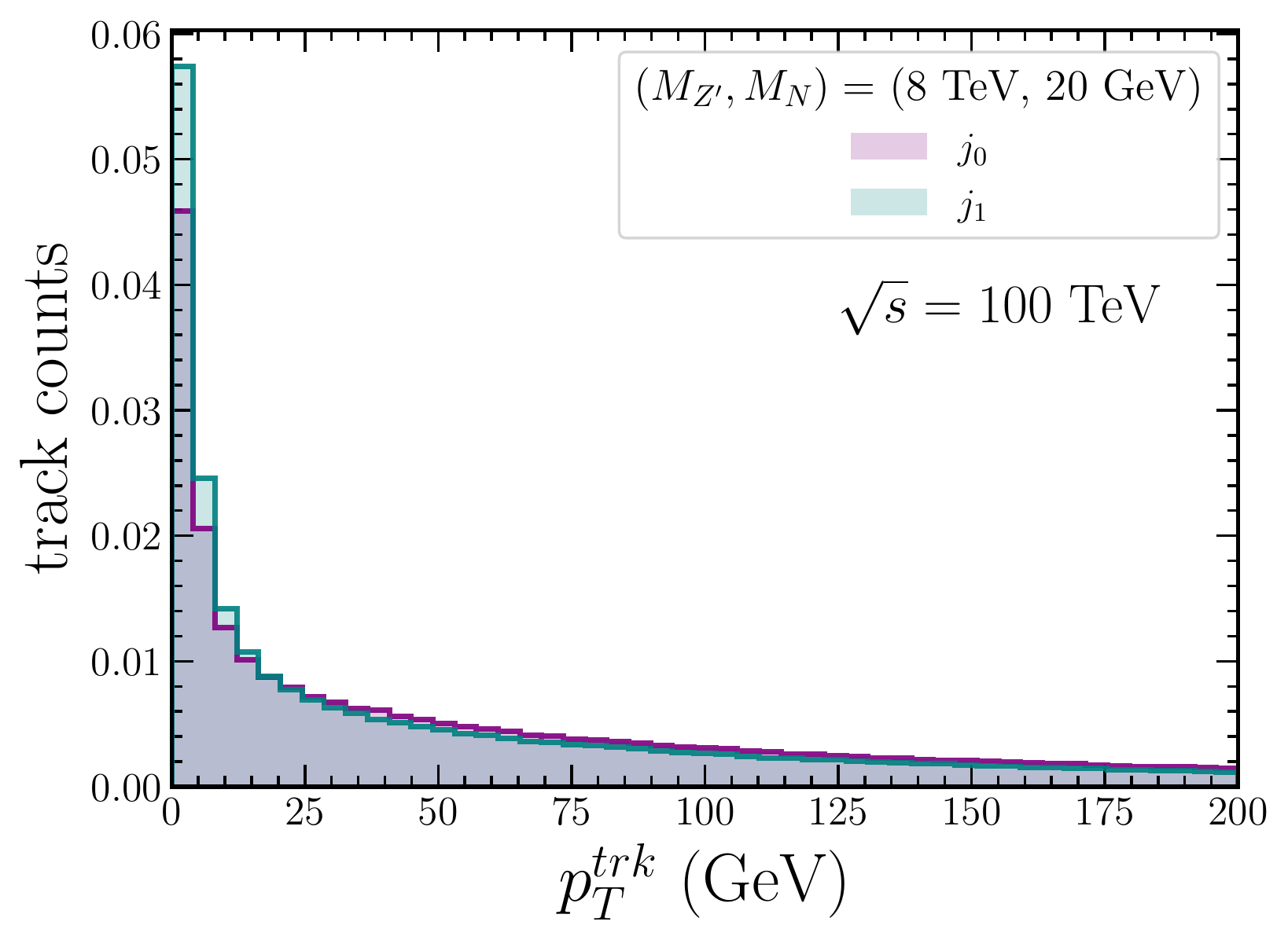}
	\includegraphics[height=0.25\textheight,width=0.45\textwidth]{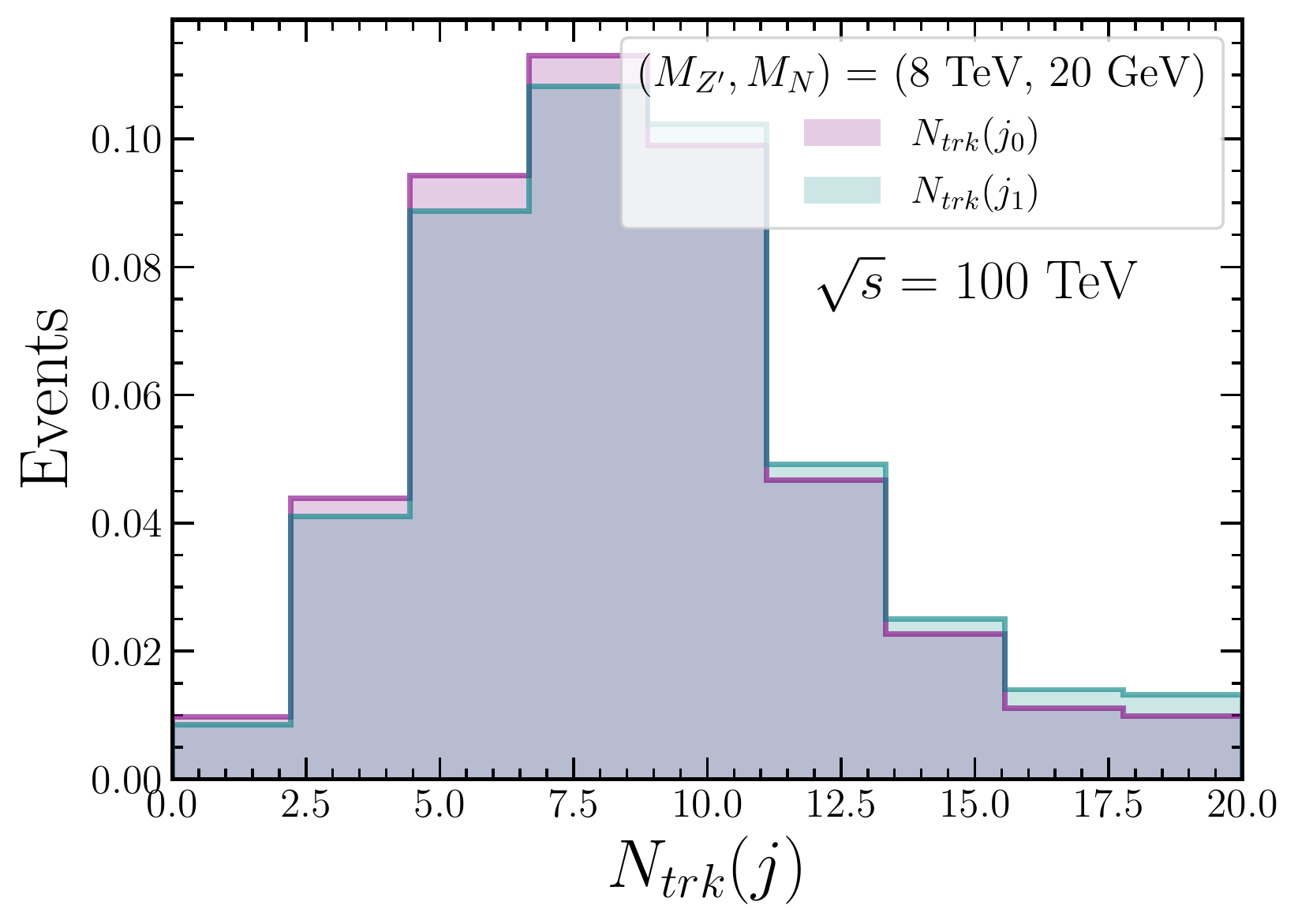}
	\caption{Left: $p_T$ of tracks associated to leading and sub-leading jets, relevant for FCC-hh. Right: Number of tracks associated to $j_{0,1}$  for FCC-hh.}
	\label{fig:ntrack}
\end{figure}
 
 \begin{figure}[t]
 	\includegraphics[height=0.28\textheight,width=0.453\textwidth]{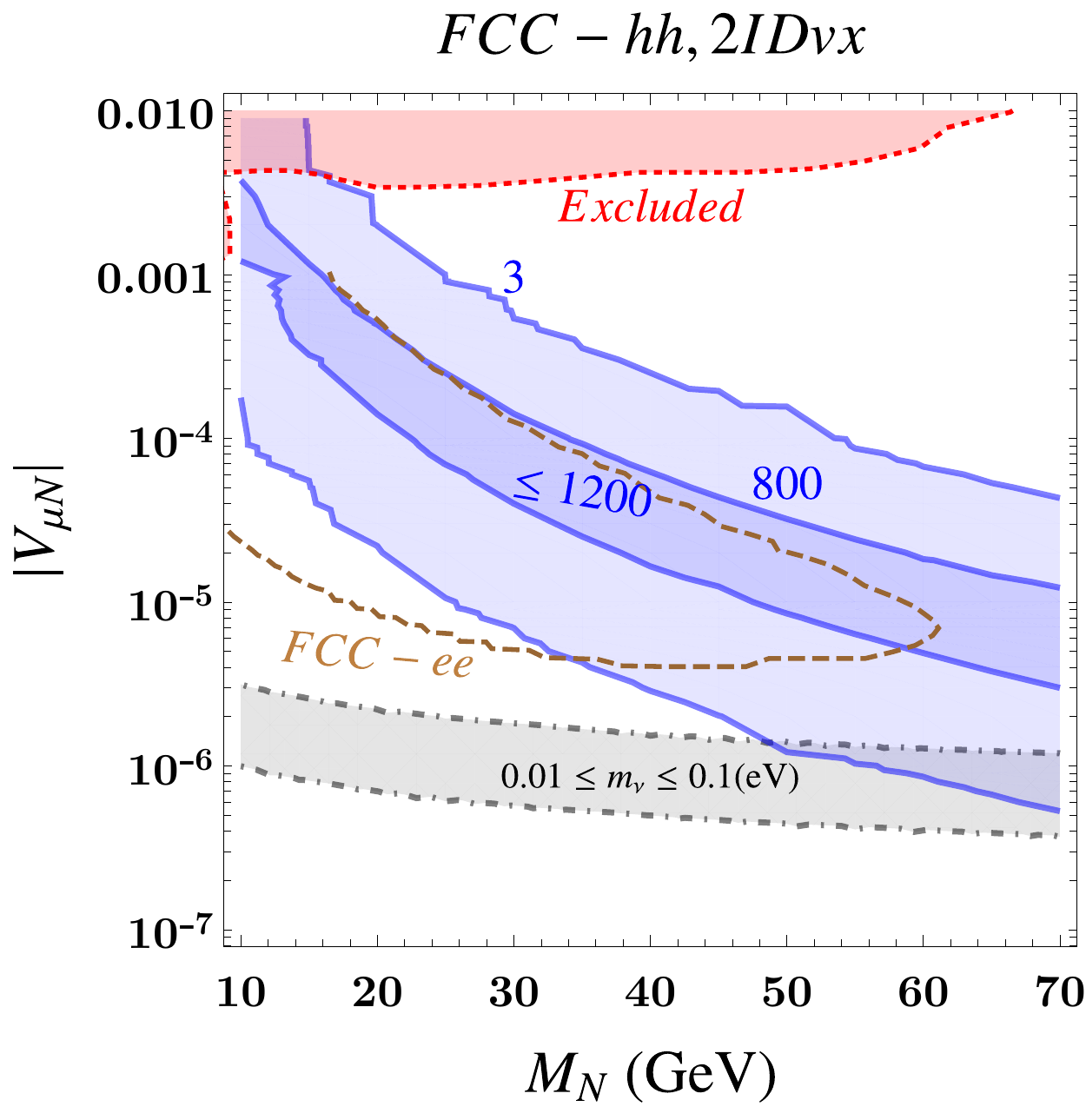}
 	\includegraphics[height=0.28\textheight,width=0.453\textwidth]{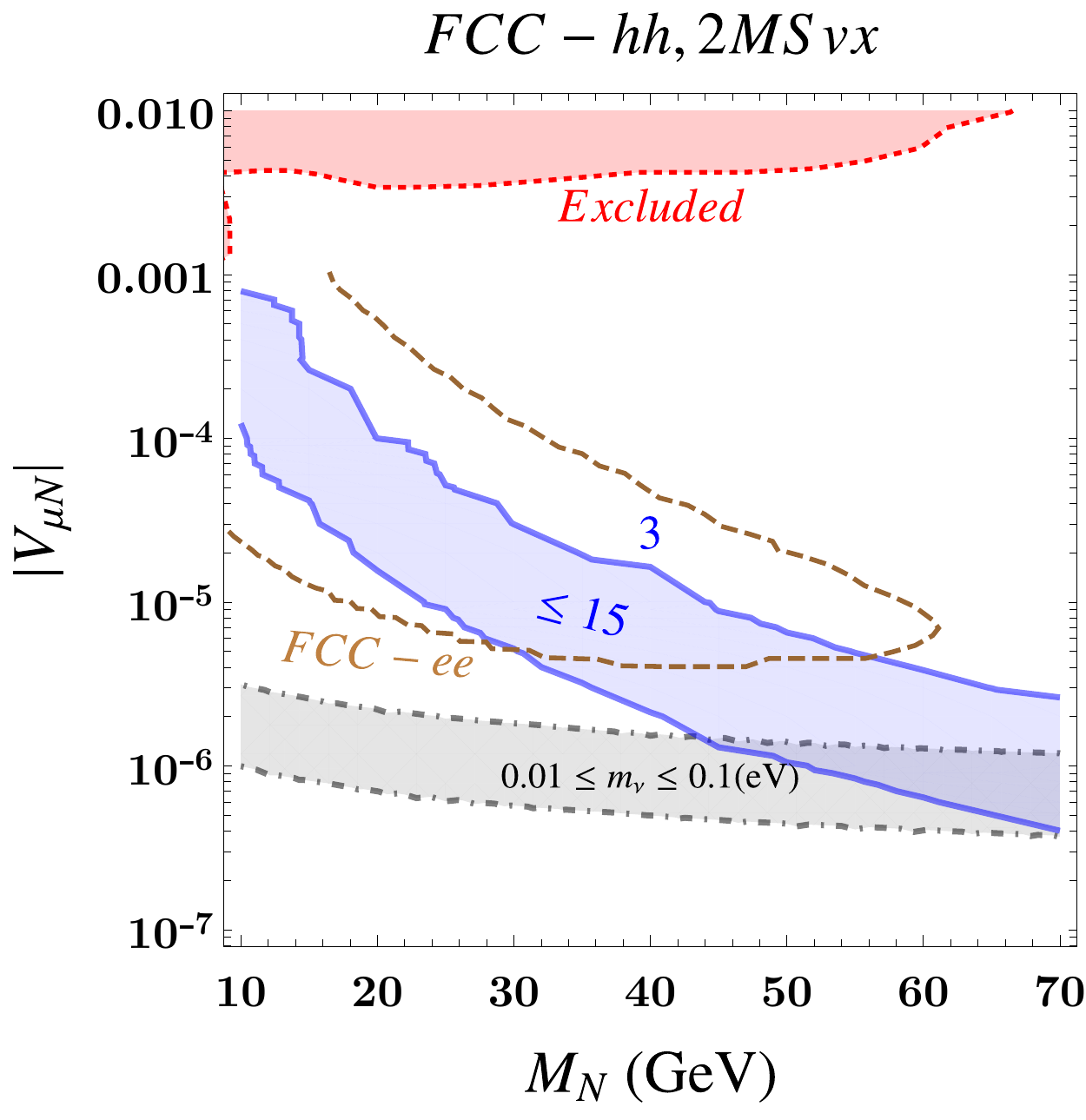}\caption{Contours of   2IDvx~(left) and 2MSvx~(right)  events for  FCC-hh with $\mathcal{L}=30\,\rm{ab}^{-1}$. The description of the brown dashed line and red shaded region remains the same as given in the previous figures. \label{fig:eventsforfcc2dis}}
 \end{figure}
 
  \begin{figure}[t]
 	\includegraphics[height=0.30\textheight,width=0.453\textwidth]{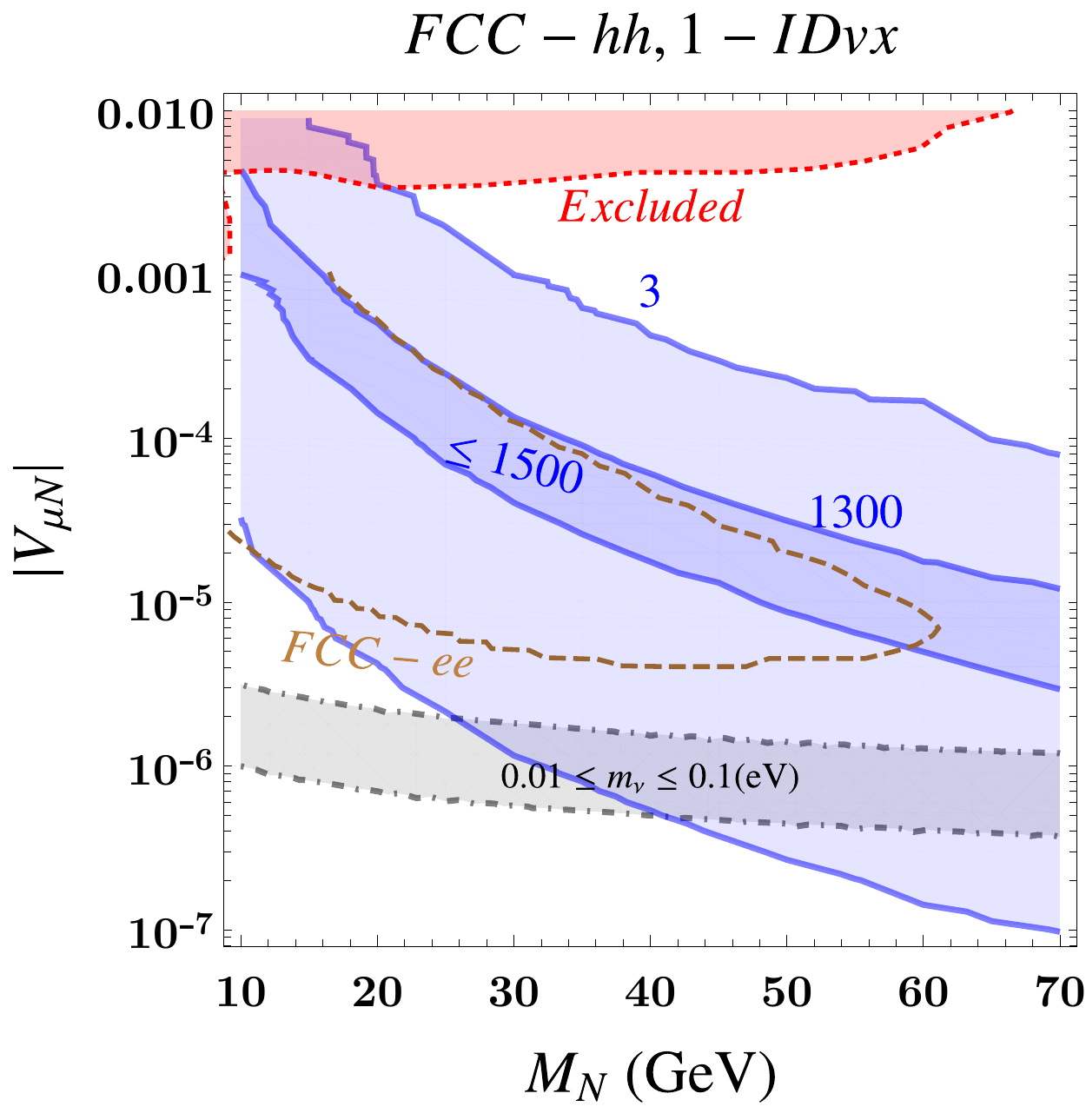}
 	\includegraphics[height=0.30\textheight,width=0.453\textwidth]{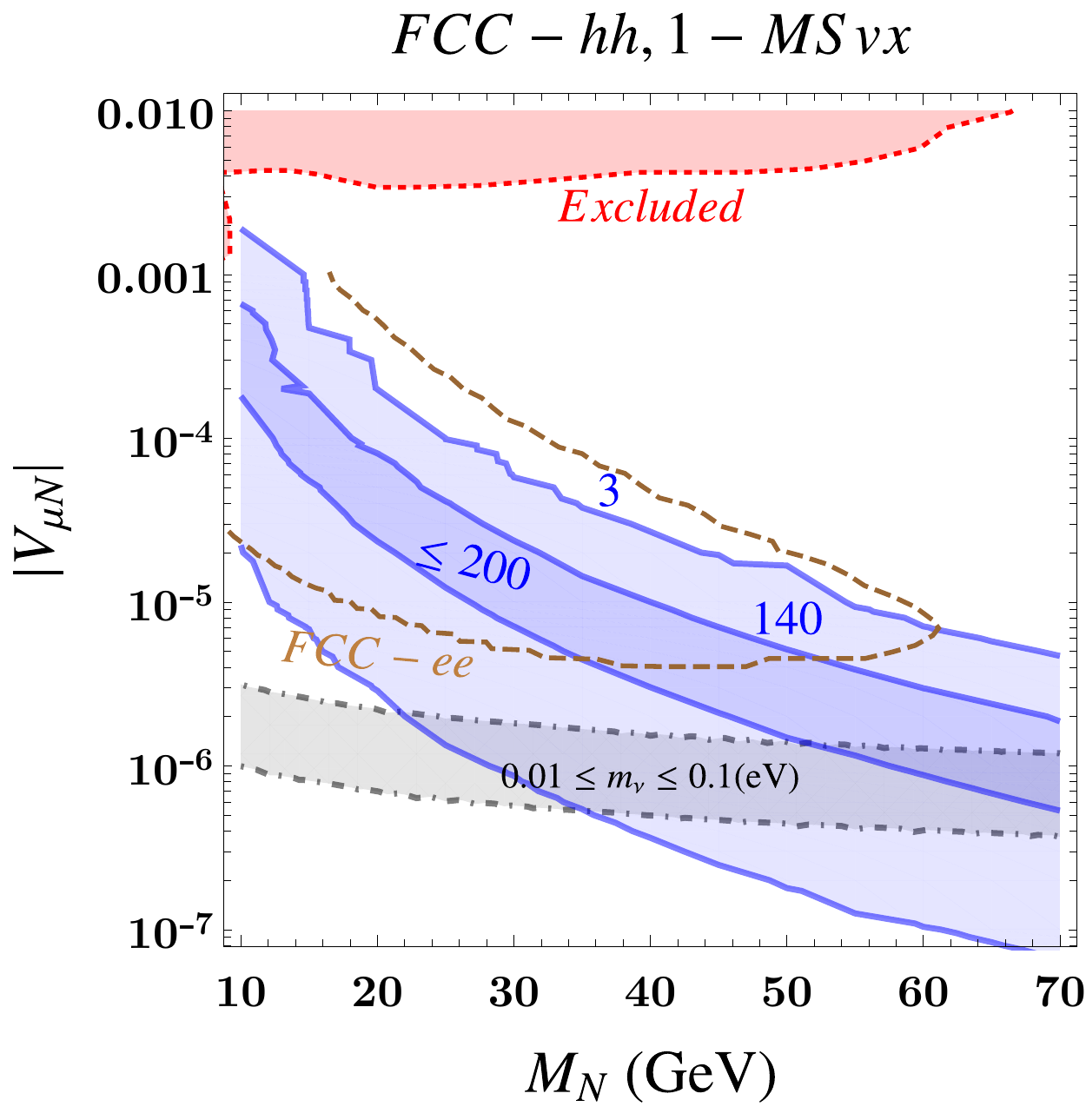}
 	\caption{Event contours for 1IDvx~(left) and 1MSvx~(right)  at FCC-hh  with $\mathcal{L}=30\,\rm{ab}^{-1}$. \label{fig:eventsforfcc1dis}}
 \end{figure}
%%%%%%%%%%%%%%%%%%%%%%%%%%%%%%%%%%%%%%%%%%%%%%%%%%%%%%%%%%%%%%%%%%%%%%%%%%%
\subsection{RHN decay signature with  IDvx:}
We consider two  RHN decaying  in the inner detector of FCC-hh, for which we use the following sets of cuts,
\begin{itemize}
\item
{\it C1:} Both the $N$s decay  within the detector length $ID_{L_1}=25$ mm, and $ID_{L_2}=1550$ mm. %We also present the analysis considering displaced decays of at least one RHN
\item
{\it C2:} The leading and sub-leading jet  $j_{0,1}$ has to satisfy  $|\eta(j)| <4.5$ and $p_T(j)\ge 300$ GeV. {A higher $p_T$ cut on the jet $p_T$ will be useful to suppress the SM backgrounds.}
\item
{\it C3:} Additionally for each of the $j_{0,1}$, number of associated tracks with $p_T\ge1 \text{GeV}$ and $|\eta|<2.5$  should satisfy $n_{trk} \ge4$. { We show the distribution of track-$p_T$ and number of tracks in Fig.~\ref{fig:ntrack} for $M_N=20$ GeV. For higher $M_N$ the number of tracks increases further.}
%(Here $n_{trk}$ is number of tracks associated with the jet.) 
\item
{\it C4:} Finally we select the events if  number of jets $n_j\ge2$ and the above-mentioned cuts are satisfied for both  the leading and sub-leading {\it fat-jets}. Additionally, similar to the analysis of HL-LHC, we also present  the results demanding displaced decay of  at least one RHN \footnote{For 1IDvx category, we tag displaced decay of one  $N$ while the other $N$ can decay anywhere.}, for which we consider only the geometric cut efficiency.  % (Both $j_0$ and $j_1$ pass IDvx cut.)
\end{itemize}
{In the left panel of Figure.~\ref{fig:eventsforfcc2dis}, we show the contours for 3 and 800 2IDvx events. We  show the event contours  corresponding to 1IDvx events  in the left panel of Fig.~\ref{fig:eventsforfcc1dis}. For both of these two scenarios, a huge number of events ${\it N_{events}} > 1000$ can be observed at FCC-hh with $\mathcal{L}=30\, \rm{ab}^{-1}$.} 
%%%%%%%%%%%%%%%%%%%%%%%%%%%%%%%%%%%%%%%%%%%%%%%%%%%%%%%%%%%%%%%%%%%%%%%%%%%
\subsection{Decay vertex in the muon spectrometer (MSvx):} 
If RHN decays  to a $\mu q q^{\prime}$ in the muon-spectrometer, we implement  the following sets of selection cuts:
\begin{itemize}
\item {\it C1:} We demand both  the RHN decay  occur within the muon spectrometer between the length $MS_{L_1}=6000$ mm, and $MS_{L_2}=9000$ mm.
\item {\it C2:} For each of the tracks originating from RHN decay vertex we impose $p_T(track)>1$ GeV , $ |\eta(track)| <2.7$ and $n_{trk} \ge4$. Additionally,  $\Sigma_{{track}} P_T(track)>60$ GeV. {These cuts are similar to the cuts that we use for  HL-LHC projection.} %\textcolor{red}{we should not say jet here. Need to discuss about this}
\item {\it C3:} Finally we select the events if two  decay vertex and the associated tracks from the vertex satisfy the above selection criterion. 
\end{itemize}
\begin{figure}[t]
	\includegraphics[height=0.30\textheight,width=0.453\textwidth]{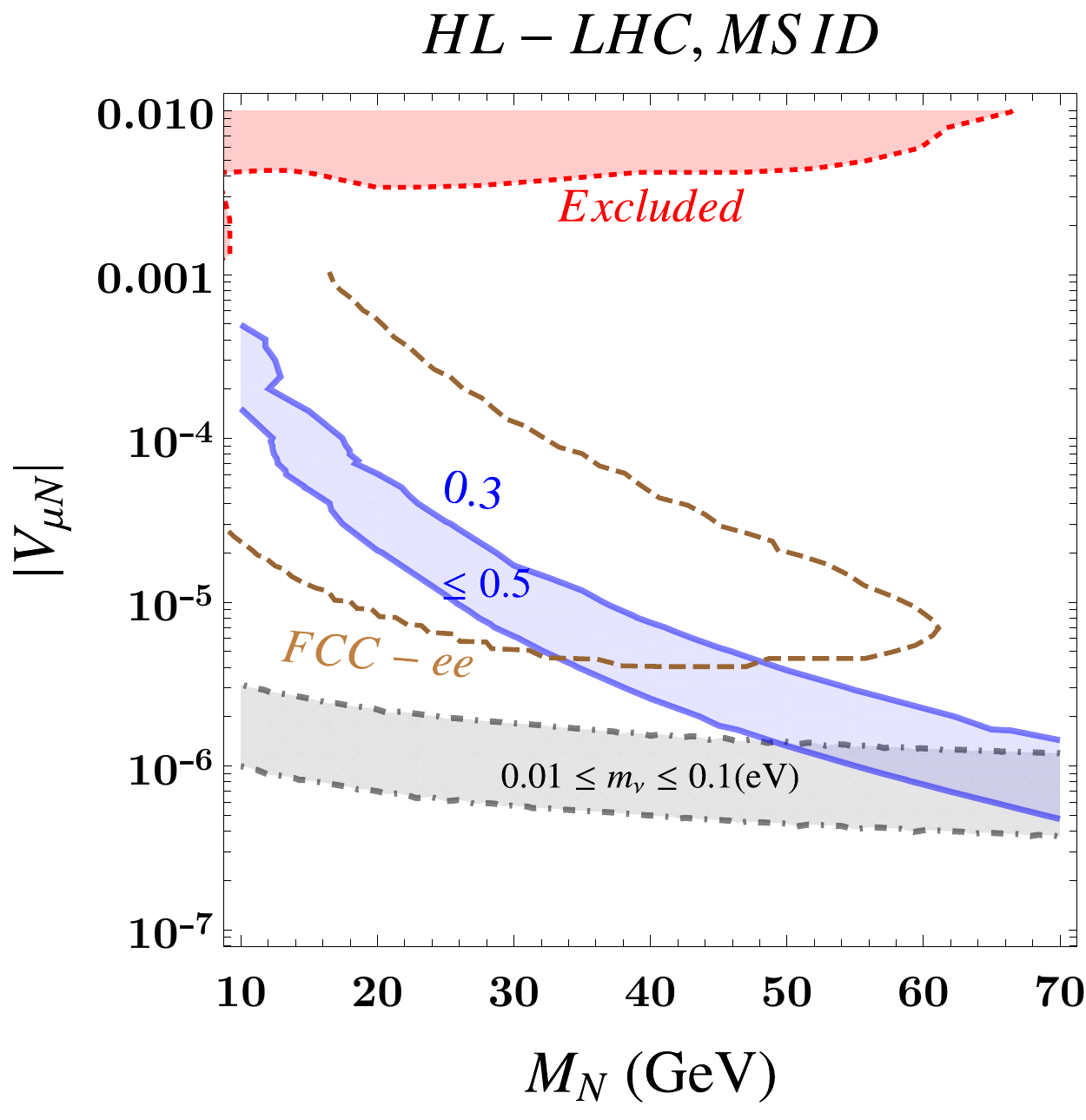} 
	\includegraphics[height=0.30\textheight,width=0.453\textwidth]{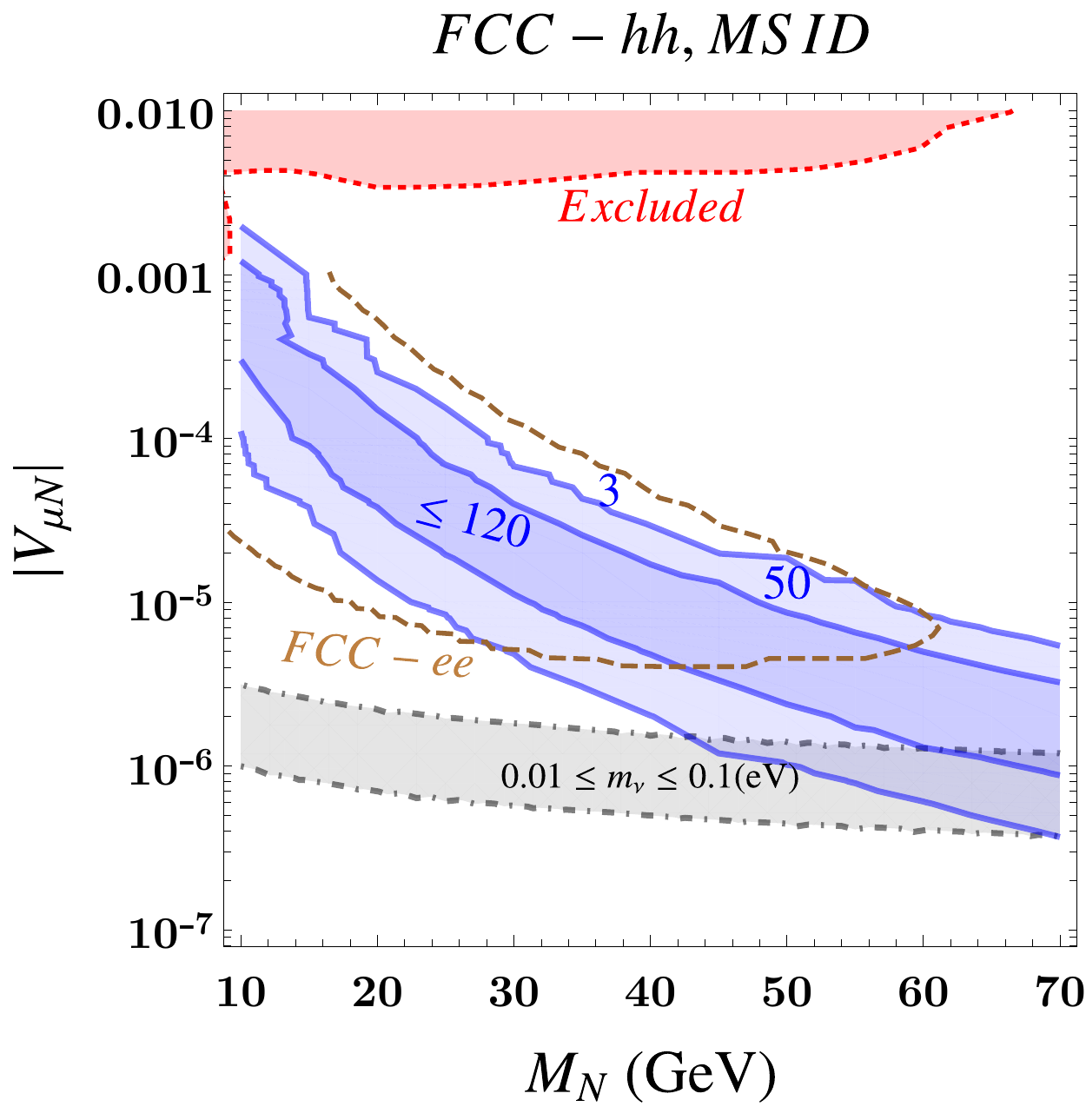} 
	\caption{Left: Number of events  for 1IDvx-1MSvx events at HL-LHC  with $\mathcal{L}=3\,\rm{ab}^{-1}$ for $M_{Z^\prime}=1$ TeV, $g^\prime=0.003$. Right: The same for FCC-hh with $\mathcal{L}=30\,\rm{ab}^{-1}$ for $M_{Z^\prime}=8 \ \rm{TeV}, \ g^\prime=0.1$.
	\label{fig:eventsforidms}}
	\end{figure}
In the right panel of Figure.~\ref{fig:eventsforfcc2dis}, we show the  event contours  for two $N$ decaying inside the muon spectrometer. We also show the number of events demanding decay of at least one $N$ in the muon spectrometer in the right panel of Fig.~\ref{fig:eventsforfcc1dis}, where we take into account  only the geometric cut-efficiency. The detection prospect of displaced $N$ decaying in the muon spectrometer is significantly larger for FCC-hh, compared to HL-LHC. This increase in ${\it N_{events}}$ occurs due to an increase in  the luminosity, and a marginal increase in the cross-section. The kinematic cut-efficiencies are more than $98\%$ for both HL-LHC and FCC-hh MSvx events, and the geometric cut-efficiencies are also somewhat similar (see Fig.~\ref{prob1} and Fig.~\ref{prob2}). Hence, these have  little effects in determining  increase in the  number of events. We also present results for the MSID event category,  where  one $N$ decays in the ID and the other one decays in the MS. The results are shown in Fig.~\ref{fig:eventsforidms} both of HL-LHC (left panel) and FCC-hh (right panel). A large number of events can be observed at FCC-hh, compared to HL-LHC, with a maximum number of events  ${\it N_{events}=120}$ for FCC-hh.

\begin{figure}
		\includegraphics[height=0.30\textheight,width=0.45\textwidth]{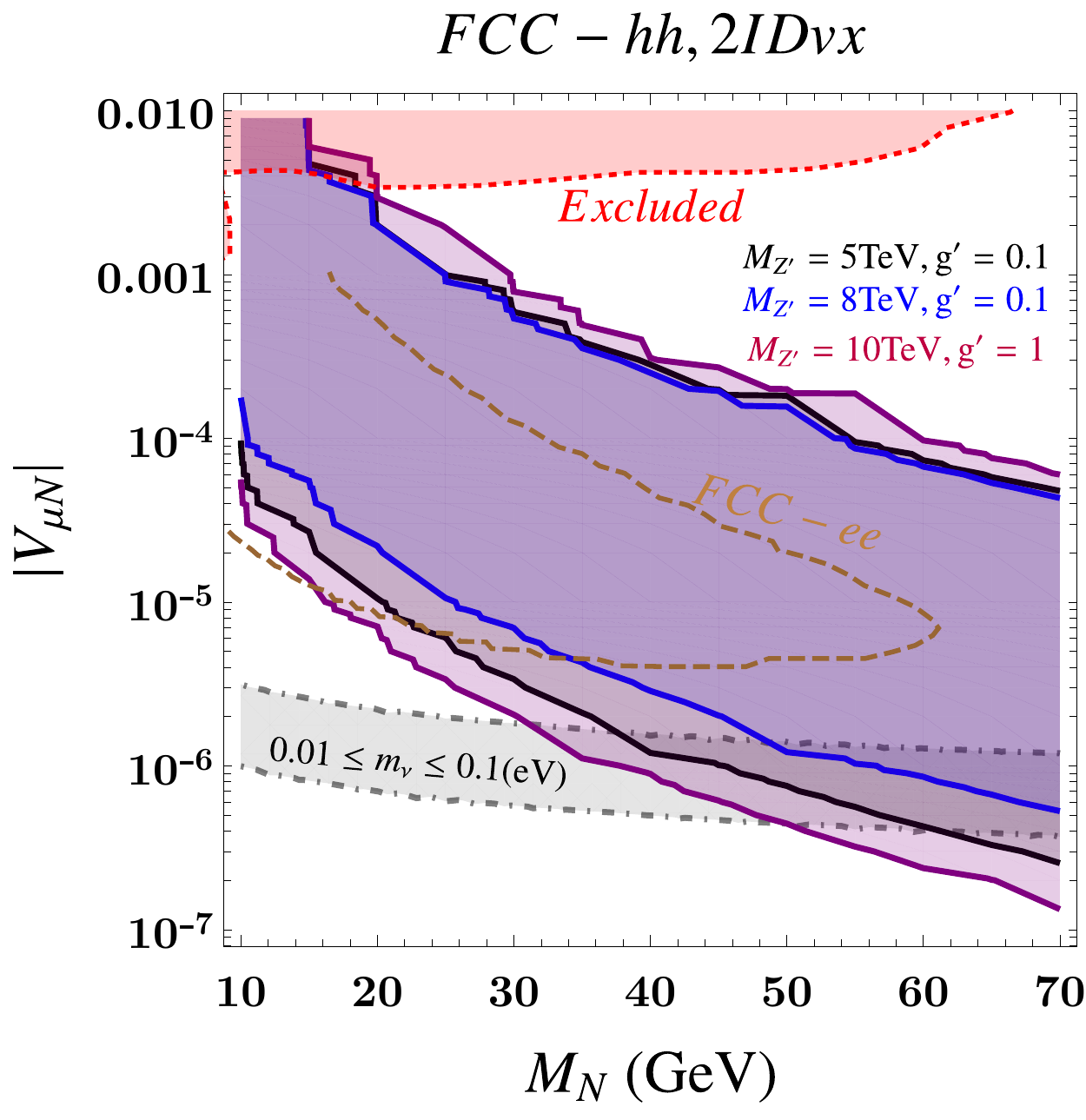}
		\includegraphics[height=0.30\textheight,width=0.45\textwidth]{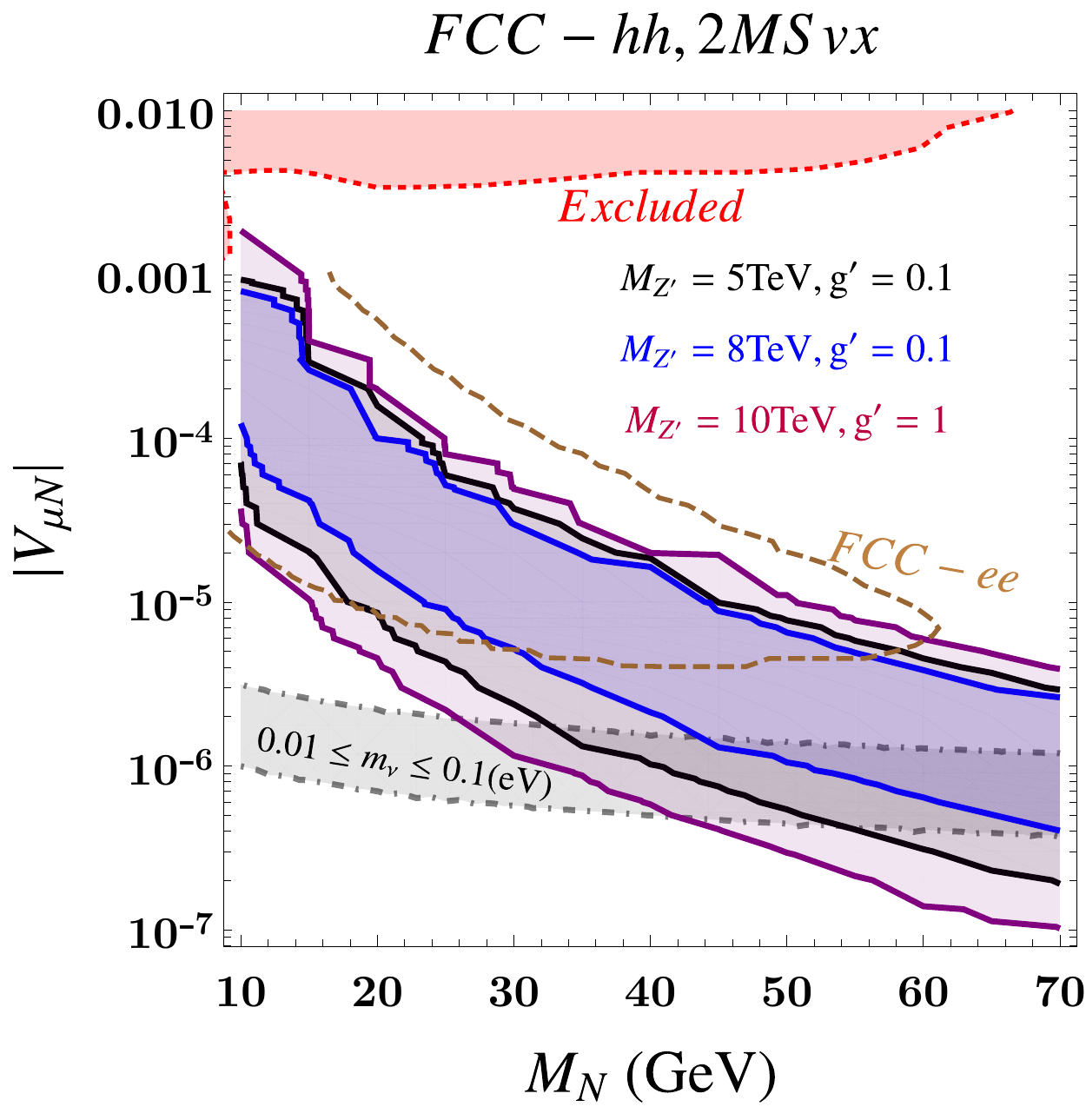} 
		\caption{ Contours for 3 events for various banchmark at FCC-hh\label{fig:diffMZline}}
	\end{figure}

{\it Sentivity reach of $|V_{\mu N}|$}: Finally, in Fig.~\ref{fig:diffMZline}, we present the sensitivity reach of $|V_{\mu N}|$ at FCC-hh  for different choices of $M_{Z^{\prime}}$ and $g^{\prime}$. We consider few benchmark points with masses $M_{Z^{\prime}}=5, 8$ TeV and coupling $g^{\prime}=0.1$, as well as, a relatively higher mass $M_{Z^{\prime}}=10$ TeV with coupling $g^{\prime}=1$. Since our signal contains sufficiently large displaced vertices, hence the SM background polluting the signal would be rather small. In the absence of any background, which we consider in this analysis,  following a Poisson distribution  only $N_{event} = 3.09$ is required at 95$\%$ confidence level. The variation of the sensitivity reach of $|V_{\mu N}|$ w.r.t  different choices of $M_N$ has been shown for 2IDvx events (left panel), and for 2MSvx events (right panel). We note that, fixing $M_{Z^{\prime}}$ and $g^{\prime}$, overall a smaller value of  active-sterile mixing $|V_{\mu N}|$ is required to obtain $N_{event} \sim 3.00$ events for the 2MSvx events, compared to 2IDvx events. This occurs  due to a larger decay length of $N$ for its decay into a muon spectrometer. Although the sensitivity reach for $|V_{\mu N}|$ for  2MSvx events have a significant overlap with the sensitivity reach for 2IDvx, however the former has added benefit as 2MSvx events in the muon spectrometer will serve as background free distinctive signal of the model.

%%%%%%%%%%%%%%%%%%%%%%%%%%%%%%%%%%%%%%%
\section{Conclusion and outlook \label{conclu}}
%%%%%%%%%%%%%%%%%%%%%%%%%%%%%%%%%%%%%%%%%%%%%%%%%%%%%%%%%%%%%%%%%%%%%%%%%%%
{The RHN $N$ in the gauged $B-L$ model can be pair-produced at a $p p$ machine via the  heavy $Z^{\prime}$ mediated channel, and give distinctive signatures with displaced vertices. The pair-production of $N$ is independent of active-sterile mixing, instead depends on the $B-L$ gauge coupling $g^{\prime}$,  mass of $Z^{\prime}$, and mass $M_N$ of $N$. Hence, even with an active-sterile mixing which is suppressed due to light neutrino mass constraint, fairly large cross-section $\sigma \sim $ fb and even higher can be achieved. We consider a relatively light $N$ with $M_N<M_W$, that satisfies light neutrino mass 
	constraint $m_{\nu}<0.1$ eV. For our choice of $M_N$, the constraint from eV light neutrino mass results in a suppressed active-sterile mixing $|V_{\ell N}| \sim 10^{-6}$. For the considered  range of $M_N$ and  small $|V_{lN}|$, the RHN undergoes three body decays with the decay vertex considerably displaced from its production vertex. We find that, $N$ undergoes a large displacement $\ell \sim mm$ upto $meter$  if  $M_N$ varies  in  between  10-70 GeV. Additionally, due to $M_N \ll M_{Z^{\prime}}$, the produced $N$ is significantly boosted, resulting in collimated decay products. }

{We analyse the displaced decay signatures of such a  light $N$ that can be probed at the high-luminosity run of the LHC (HL-LHC) with c.m.energy $\sqrt{s}=14$ TeV and the future $p p $ machine FCC-hh that can operate with c.m.energy $\sqrt{s}=100$ TeV. Specifically, we consider two scenarios, which are $N$ decays a) within the  inner-detector (ID) of the HL-LHC and FCC-hh detectors, and b) within the first few layers of the muon spectrometer (MS). For the former,  $N$ should have proper decay length $\ell \sim$ mm to hundreds of mm, while for later, the decay length should be $\ell \sim $ m.  We emphasize  that the  signal description between a) and b) differ widely. For a detail analysis of the signature, we specifically consider $N \to \mu q q^{\prime}$ decay mode. For a), i.e., if the decay of $N$ occurs in the ID, the collimated decay products of $N$ result in a displaced {\it fat-jet}. For this case, we thus analyse  model signature with  two displaced {\it fat-jets}, and further extend the analysis to take into account  displaced decay of at-least one $N$. We find  that $\mathcal{O}(10)$  displaced {\it fat-jet} events can be observed at HL-LHC with 3\,$\rm{ab}^{-1}$ of luminosity,   and for FCC-hh, a significantly large number of events $ \mathcal{O}(1000)$ can be observed with $\mathcal{L}=30\, \rm{ab}^{-1}$.}

{For b), i.e., if $N$ decays in the MS, {\it fat-jet} description can not be used, since energy deposit in the calorimeter,  informations about the track in the ID etc are missing, which are being used for the formation of a jet. In this case, we instead perform an analysis that relies on the  properties of tracks in the MS.  
	We apply selection cuts on the number of tracks originated from RHN decay, $p_T$ and $|\eta|$, and summation of $p_T$ of the tracks associated with $N$. We find that for HL-LHC, even after using the full integrated luminosity, the sensitivity reach is significantly low. For FCC-hh, this  improves by order of magnitude, as a large number of events $\mathcal{O}(100)$ can be observed with the full integrated luminosity. We also extend the analysis for the case, when one of the $N$ decays in the ID and the other $N$ decays in the MS, for which results are similar to b).}

%\label{sec:conclusion}
{Our proposed signature can potentially be explored for other decay modes of $N$, including $N \to e q q^{\prime}$, as well as   $N \to \nu q q^{\prime}$ final states. The final state with electron depends on active sterile mixing $|V_{eN}|$, and hence choosing a large $V_{eN}$ compared to $V_{\mu N}$ will make the contribution from $N \to e q q^{\prime}$ large. The $N \to \nu q q^{\prime}$ decay instead depends on all possible $|V_{lN}|$. The final state in addition to displaced {\it fat-jet} or charged tracks, will also carry missing transverse energy. By applying a veto on MET will hence reduce the contamination from $\nu q q^{\prime}$ decay mode. A detail analysis of the model signature including both $e q q^{\prime}$ and $\nu q q^{\prime}$ decays will be presented elsewhere.}
%%%%%%%%%%%%%%%%%%%%%%%%%%%%%%%%%%%%%%%%%%%%%%%%%%%%%%%%%%%%%%%%%%%%%%%%%%%%%%%%%
\acknowledgments
%%%%%%%%%%%%%%%%%%%%%%%%%%%%%%%%%%%%%%%%%%%%%%%%%%%%%%%%%%%%%%%%%%%%%%%%%%%%%%%%%
SK is supported by Elise-Richter grant project number V592-N27 and FFD by a UK STFC consolidated grant (Reference ST/P00072X/1). We thank F. Blekman, S. Pagan Griso for several useful discussions.
RP acknowledge SAMKHYA: High-Performance Computing Facility provided by the Institute of Physics (IoP), Bhubaneswar. The authors thank Dr. Benjamin Radburn-Smith and Dr. Aruna Nayak for useful discussions regarding CMS analysis.
%%%%%%%%%%%%%%%%%%%%%%%%%%%%%%%%%%%%%%%%%%%%%%%%%%%%%%%%%%%%%%%%%%%%%%%%%%%%%%%%%
\bibliographystyle{JHEP}
\bibliography{displaced_fat_jet_RP}
\end{document}